\documentclass[iop,apj,numberedappendix,appendixfloats]{emulateapj}

\newcommand{\myemail}{adelmoro@mpe.mpg.de}
\usepackage{times}
\usepackage{epsfig} 
\usepackage{amssymb}
\usepackage{graphics}
\usepackage{natbib} 

\newcommand{\cgs}{erg~cm$^{-2}$~s$^{-1}$}  
\newcommand{\ergs}{erg~s$^{-1}$} 
\newcommand{\nh}{$N_{\rm H}$} 
\newcommand{\ch}{{\it Chandra}}
\newcommand{\xmmn}{{\it XMM-Newton}}
\newcommand{\xmm}{{\it XMM}}

\newcommand{\nus}{{\it NuSTAR}} 

\shorttitle{Average broad-band X-ray spectra with NuSTAR}
\shortauthors{A.~Del Moro et al.}

\begin{document}

\title{The \nus\ Extragalactic Survey: Average broad-band X-ray spectral properties of the \nus\ detected AGN}

%% Use \author, \affil, and the \and command to format
%% author and affiliation information.
%% Note that \email has replaced the old \authoremail command
%% from AASTeX v4.0. You can use \email to mark an email address
%% anywhere in the paper, not just in the front matter.
%% As in the title, use \\ to force line breaks.

\author{A.~Del~Moro\altaffilmark{1,2}, D.~M.~Alexander\altaffilmark{2}, J.~A.~Aird\altaffilmark{3}, F.~E.~Bauer\altaffilmark{4,5,6}, F.~Civano\altaffilmark{7,8}, J.~R.~Mullaney\altaffilmark{9}, D.~R.~Ballantyne\altaffilmark{10}, W.~N.~Brandt\altaffilmark{11,12,13}, A.~Comastri\altaffilmark{14}, P.~Gandhi\altaffilmark{15}, F.~A.~Harrison\altaffilmark{16}, G.~B.~Lansbury\altaffilmark{2,3}, L.~Lanz\altaffilmark{7}, B.~Luo\altaffilmark{17}, S.~Marchesi\altaffilmark{18}, S.~Puccetti\altaffilmark{19,20}, C.~Ricci\altaffilmark{4,21,22}, C.~Saez\altaffilmark{23}, D.~Stern\altaffilmark{24},  E.~Treister\altaffilmark{4} and L.~Zappacosta\altaffilmark{19} 
}
\affil{$^1$ Max-Planck-Institut f{\"u}r Extraterrestrische Physik (MPE), Giessenbachstrasse 1,85748 Garching, Germany; email: \myemail}
\affil{$^2$ Centre for Extragalactic Astronomy, Department of Physics, Durham University, South Road, Durham, DH1 3LE, UK}
\affil{$^3$ Institute of Astronomy, University of Cambridge, Madingley Road, Cambridge, CB3 0HA, UK}
\affil{$^4$ Instituto de Astrof\'{\i}sica and Centro de Astroingenier{\'{\i}}a, Facultad de F{\'{i}}sica, Pontificia Universidad Cat{\'{o}}lica de Chile, Casilla 306, Santiago 22, Chile}
\affil{$^5$ Millennium Institute of Astrophysics (MAS), Nuncio Monse{\~{n}}or S{\'{o}}tero Sanz 100, Providencia, Santiago, Chile}
\affil{$^6$ Space Science Institute, 4750 Walnut Street, Suite 205, Boulder, Colorado 80301}
\affil{$^7$ Department of Physics and Astronomy, Dartmouth College, 6127 Wilder Laboratory, Hanover, NH 03755, USA}
\affil{$^8$ Harvard-Smithsonian Center for Astrophysics, 60 Garden Street, Cambridge, MA 02138, USA}
\affil{$^9$ Department of Physics \& Astronomy, University of Sheffield, Sheffield, S3 7RH, UK}
\affil{$^{10}$ Center for Relativistic Astrophysics, School of Physics, Georgia Institute of Technology, Atlanta, GA 30332, USA}
\affil{$^{11}$ Department of Astronomy and Astrophysics, 525 Davey Lab, The Pennsylvania State University, University Park, PA 16802, USA}
\affil{$^{12}$ Institute for Gravitation and the Cosmos, The Pennsylvania State University, University Park, PA 16802, USA}
\affil{$^{13}$ { Department of Physics, 104 Davey Lab, The Pennsylvania State University, University Park, PA 16802, USA}}
\affil{$^{14}$ INAF - Osservatorio Astronomico di Bologna, { via Gobetti 93/3, 40129 Bologna}, Italy}
\affil{$^{15}$ Department of Physics \& Astronomy, Faculty of Physical Sciences and Engineering, University of Southampton, Southampton, SO17 1BJ, UK}
\affil{$^{16}$ Cahill Center for Astrophysics, 1216 East California Boulevard, California Institute of Technology, Pasadena, CA 91125, USA} 
\affil{$^{17}$ School of Astronomy and Space Science, Nanjing University, Nanjing 210093, China}
\affil{$^{18}$ Department of Physics \& Astronomy, Clemson University, Clemson, SC 29634, USA}
\affil{$^{19}$ INAF-Osservatorio Astronomico di Roma, via Frascati 33, I-00040 Monteporzio Catone, Italy}
\affil{$^{20}$ { Agenzia Spaziale Italiana-Unit{\`a} di Ricerca Scientifica, Via del Politecnico, 00133 Roma, Italy}}
\affil{$^{21}$ EMBIGGEN Anillo, Concepcion, Chile}
\affil{$^{22}$ Kavli Institute for Astronomy and Astrophysics, Peking University, Beijing 100871, China}
\affil{$^{23}$ { Observatorio Astron\'omico Cerro Cal\'an, Departamento de Astronom\'ia, Universidad de Chile, Casilla 36-D, Santiago, Chile}}
\affil{$^{24}$ Jet Propulsion Laboratory, California Institute of Technology, 4800 Oak Grove Drive, Mail Stop 169-221, Pasadena, CA 91109, USA}

\begin{abstract}
We present a study of the average X-ray spectral properties of the sources detected by the \nus\ extragalactic survey, comprising observations of the Extended-\ch\ Deep Field South (E-CDFS), Extended Groth Strip (EGS) and the Cosmic Evolution Survey (COSMOS). The sample includes 182 \nus\ sources (64 detected at $8-24$~keV), with $3-24$~keV fluxes ranging between $f_{\rm 3-24~keV}\approx10^{-14}$ and $6\times10^{-13}$~\cgs\ ($f_{\rm 8-24~keV}\approx3\times10^{-14}-3\times10^{-13}$~\cgs) and redshifts in the range of $z=0.04-3.21$. We produce composite spectra from the \ch\ $+$ \nus\ data ($E\approx2-40$~keV, rest frame) for all the sources with redshift identifications (95\%) and investigate the intrinsic, average spectra of the sources, divided into broad-line (BL) and narrow-line (NL) AGN, and also in different bins of X-ray column density and luminosity. The average power-law photon index for the whole sample is $\Gamma=1.65_{-0.03}^{+0.03}$, flatter than the $\Gamma\approx1.8$ typically found for AGN. While the spectral slope of BL and X-ray unabsorbed AGN is consistent with the typical values { (${\Gamma=1.79_{-0.01}^{+0.01}}$)}, a significant flattening is seen in NL AGN and heavily absorbed sources ($\Gamma=1.60_{-0.05}^{+0.08}$ and ${\Gamma=1.38_{-0.12}^{+0.12}}$, respectively), likely due to the effect of absorption and to the contribution from the Compton-reflection component to the high energy flux ($E>10$~keV). We find that the typical reflection fraction in our spectra is $R\approx0.5$ (for $\Gamma=1.8$), with a tentative indication of an increase of the reflection strength with X-ray column density. While there is no { significant} evidence for a dependence of the photon index with X-ray luminosity in our sample, we find that ${R}$ decreases with luminosity, with relatively high levels of reflection (${R\approx1.2}$) for $L_{\rm 10-40~keV}<10^{44}$~erg~s$^{-1}$ and ${R\approx0.3}$ for $L_{\rm 10-40~keV}>10^{44}$~erg~s$^{-1}$ AGN{, assuming a fixed spectral slope of $\Gamma=1.8$}.  
\end{abstract}


\section{Introduction}

Studies of the cosmic X-ray background (CXB) have demonstrated that the diffuse X-ray emission observed as a background radiation in X-ray surveys can be explained by the summed emission from unresolved X-ray sources, mainly active galactic nuclei (AGN) at low and high redshift. Moreover, the majority of these AGN must be obscured to reproduce the characteristic CXB spectrum peak at $E\approx20-30$~keV (\citealt{comastri1995, treister2005,gilli2007, treister2009, ballantyne2009}). In particular, synthesis models of the CXB require a population of heavily obscured AGN, defined as Compton thick (CT), where the equivalent hydrogen column density (\nh) exceeds the inverse of the Thomson scattering cross-section (\nh$>1/\sigma_{\rm T}\approx1.5\times10^{24}$~cm$^{-2}$). The fraction of such sources and their space density, however, are still uncertain and vary from model to model, depending on different parameter assumptions ($\sim$10-30\%, e.g.,\citealt{gilli2007,treister2009,akylas2012}). The main differences between these models reside in the adopted \nh\ distribution of the AGN population, the X-ray luminosity function (XLF) and the AGN spectral models. Many of these parameters are degenerate and this prevents us from securely determining the composition of the CXB at its peak (e.g. \citealt{treister2009,akylas2012}).

At energies $E<10$~keV the sensitive surveys undertaken with \ch\ and \xmmn\ have allowed us to resolve directly up to 90\% of the CXB as individual sources, placing important constraints on the total AGN population (\citealt{hickox2006, xue2012}). However, even these surveys struggle to detect and identify the most-obscured AGN, or tend to underestimate the intrinsic column density of these sources (e.g. \citealt{delmoro2014, lansbury2014, lansbury2015}), especially at high redshift, leaving significant uncertainties on the intrinsic \nh\ distribution. Moreover, the lack of direct sensitive measurements of the AGN population at high energies ($E\gtrsim10$~keV), due to the lack of sensitive hard X-ray telescopes until the past few years, { has} only allowed us to resolve directly $\sim1-2$\% of the CXB at its peak (e.g., with {\it Swift-BAT} or {\it INTEGRAL}; \citealt{krivonos2007, ajello2008, bottacini2012}). Therefore, our knowledge and models of the CXB composition at high energies solely rely on extrapolations from lower energies.  

The extragalactic survey program undertaken by \nus, the first sensitive hard X-ray telescope ($E\approx3-79$~keV) with focussing optics (\citealt{harrisonf2013}), provides great improvements on our understanding of the AGN population at $E>10$~keV. With the sources detected by \nus\ in the Extended \ch\ Deep Field-South (E-CDFS; \citealt{mullaney2015b}), Cosmic Evolution Survey (COSMOS; \citealt{civano2015}), Extended Groth Strip (EGS; Del Moro et al., in preparation) and the serendipitous survey fields (\citealt{alexander2013}; \citealt{lansbury2017}) we are now able to resolve directly $\approx$35\% of the CXB at $E=8-24$~keV (\citealt{harrisonf2016}), a much higher fraction than possible with pre-\nus\ telescopes. However, the first studies of the XLF with \nus\ (\citealt{aird2015b}) have shown that there are still degeneracies in the models to reconcile the XLF derived from \nus\ sources with extrapolations from the lower energies (2-10~keV) XLF, in particular related to the distribution of absorbing column densities and the intrinsic spectral properties of AGN, such as the strength of the Compton-reflection component. Detailed X-ray spectral analysis of the \nus\ sources is required to break these degeneracies and place tighter constraints on the measurements of the XLF and the AGN population contributing to the CXB. 

In this paper we aim to investigate the average broad-band X-ray ($\sim0.5-25$~keV) spectral properties of the \nus\ sources detected in the E-CDFS, COSMOS and EGS fields, in order to constrain the intrinsic spectral properties of the sources and measure the typical strength of the Compton reflection. To this end we produce rest-frame composite spectra at $\sim2-40$~keV (rest frame) with \ch$+$\nus\ data for the whole sample and for various subsamples, to investigate how the spectral parameters might vary between broad-line (BL) and narrow-line (NL) AGN, or as a function of X-ray column density and luminosity. A study focussing on the spectral analysis of the brightest hard-band ($8-24$~keV) detected sources is presented in a companion paper (Zappacosta et al. 2017, in prep).

Throughout the paper we assume a cosmological model with $H_0=70\ \rm km~s^{-1}~Mpc^{-1}$, $\Omega_M=0.27$ and $\Omega_{\Lambda}=0.73$ \citep{spergel2003}. All the errors and upper limits are quoted at a 90\% confidence level, unless otherwise specified. 

\section{Data and catalogues}\label{data}

The \nus\ extragalactic survey program consists of tiered observations of well-known survey fields: (i) a set of deep tiled pointings covering the E-CDFS (\citealt{mullaney2015b}) and EGS (Del Moro et al. in prep) fields, with an area of $\approx30\times30$ and $\approx12\times54$~arcmin$^2$, respectively, and a total exposure of 1.49~Ms in each of the two fields, reaching a maximum exposure of $\approx$220~ks in E-CDFS, and $\approx$280~ks in EGS (at $3-24$~keV in each focal plane module, FPM, vignetting-corrected); (ii) a medium-depth set of 121 tiled pointings covering $\approx$1.7~deg$^2$ of the COSMOS field \citep{civano2015} for a total exposure of 3.12~Ms, with a maximum depth of $\approx$90~ks in the central $\sim$1.2~deg$^2$; (iii) a serendipitous survey consisting of all serendipitous sources detected in any \nus\ targeted observation (excluding all sources associated with the target; \citealt{alexander2013}; \citealt{lansbury2017}). This latter tier of the survey spans a wide range of depths (see e.g., \citealt{aird2015b, lansbury2017}) and has the largest sky coverage, reaching $\approx13$~deg$^2$ to date (still ongoing).  

Since in this work we require low-energy ($E<10$~keV) \ch\ data, together with the \nus\ data, to produce broad X-ray band composite spectra, we limit our analyses to the sources detected in the E-CDFS, EGS and COSMOS fields, which have good \ch\ coverage and redshift completeness (spectroscopic and photometric), while we exclude the serendipitous survey sources due to the heterogeneity of the available ancillary data. Although \xmmn\ observations are also available for the E-CDFS and COSMOS fields, we do not include these data in our analyses, because combining the data from different X-ray instruments to produce the composite spectra can cause significant distortions in the resulting spectra (see Sect. \ref{ave}). Our sample consists of 182 AGN (see Table \ref{tab.1}), 49 detected in E-CDFS \citep{mullaney2015b}, 42 in EGS (Del Moro et al., in prep) and 91 detected in COSMOS \citep{civano2015}. Of these sources 64 ($\sim$35\%) are detected in the \nus\ hard band (HB; $8-24$ keV).\footnote{{ Most of the sources that are formally HB-undetected ($\sim$65\%) are actually detected also above 8~keV, but with lower significance than the false-probability threshold adopted by \citet{civano2015}, \citet{mullaney2015b} and Del Moro et al. (in prep).}}      
\begin{table}
\caption{\nus\ source sample summary.}
\begin{center}
\begin{tabular}{l c c c c c}
\hline
\hline
\rule[-1.5mm]{0pt}{3ex}Field & No. of Sources & HB$^a$ & Redshift$^b$ & BL AGN & NL AGN \\
\hline
\rule[-1.5mm]{0pt}{3ex}E-CDFS    & 49   &  19  & 45 { (42)} & 18 { (18)}& 19 { (19)} \\
\rule[-1.5mm]{0pt}{3ex}EGS       & 42   &  13  & 42 { (33)} & 18 { (18)} & 21 { (14)} \\ 
\rule[-1.5mm]{0pt}{3ex}COSMOS    & 91   &  32  & 86 { (80)} & 40 { (40)} & 29 { (29)}\\
\hline
\rule[-1.5mm]{0pt}{4ex}Total     & 182  &  64 & 173 { (155)} & 76 { (76)} & 69 { (62)} \\
\hline
\end{tabular}
\end{center}
Notes: $^a$ Number of hard-band (HB; $8-24$~keV) detected sources; $^b$ Including both spectroscopic and photometric redshifts; { the number of spectroscopic redshifts is reported in parenthesis (also in columns 5 and 6)}.
\label{tab.1}
\end{table}%

{ In} \citet{civano2015} and \citet{mullaney2015b}, the \nus\ sources have been matched to the \ch\ and/or \xmmn\ point-source catalogues available in these fields (\citealt{lehmer2005, brusa2010, xue2011, ranalli2013, nandra2015, civano2016, marchesi2016}) using a nearest-neighbour approach with a matching radius of 30$\arcsec$, to identify a lower X-ray energy counterpart and thus obtain the multiwavelength information, such as the spectroscopic or photometric redshift and optical classification. The same approach has been used for the sources in the EGS field (Del Moro et al., in preparation). 11 out of 49 ($\sim$22\%) sources in E-CDFS, 14 out of 91 sources ($\sim$15\%) in COSMOS and 10 out of 42 sources ($\sim$24\%) in EGS can be associated with multiple counterparts within the 30$\arcsec$ matching radius. In our spectral analyses we use the primary counterpart identified by \citet{civano2015} for the COSMOS sources, while for the E-CDFS and EGS sources, for which a primary counterpart amongst the possible candidates has not been specified in the catalogues (see, e.g., \citealt{mullaney2015b}), we choose the \ch\ counterpart that more closely matches the \nus\ flux at $3-7$~keV (see Sect. \ref{spec}). We note that the \nus\ flux for these sources could still include some contribution from the secondary counterparts, however this contribution is expected to be {limited}. We estimated that the fluxes of the secondary counterparts are typically $<$35\% of the NuSTAR flux at $3-7$~keV.
\begin{figure}
\centerline{
\includegraphics[scale=0.7]{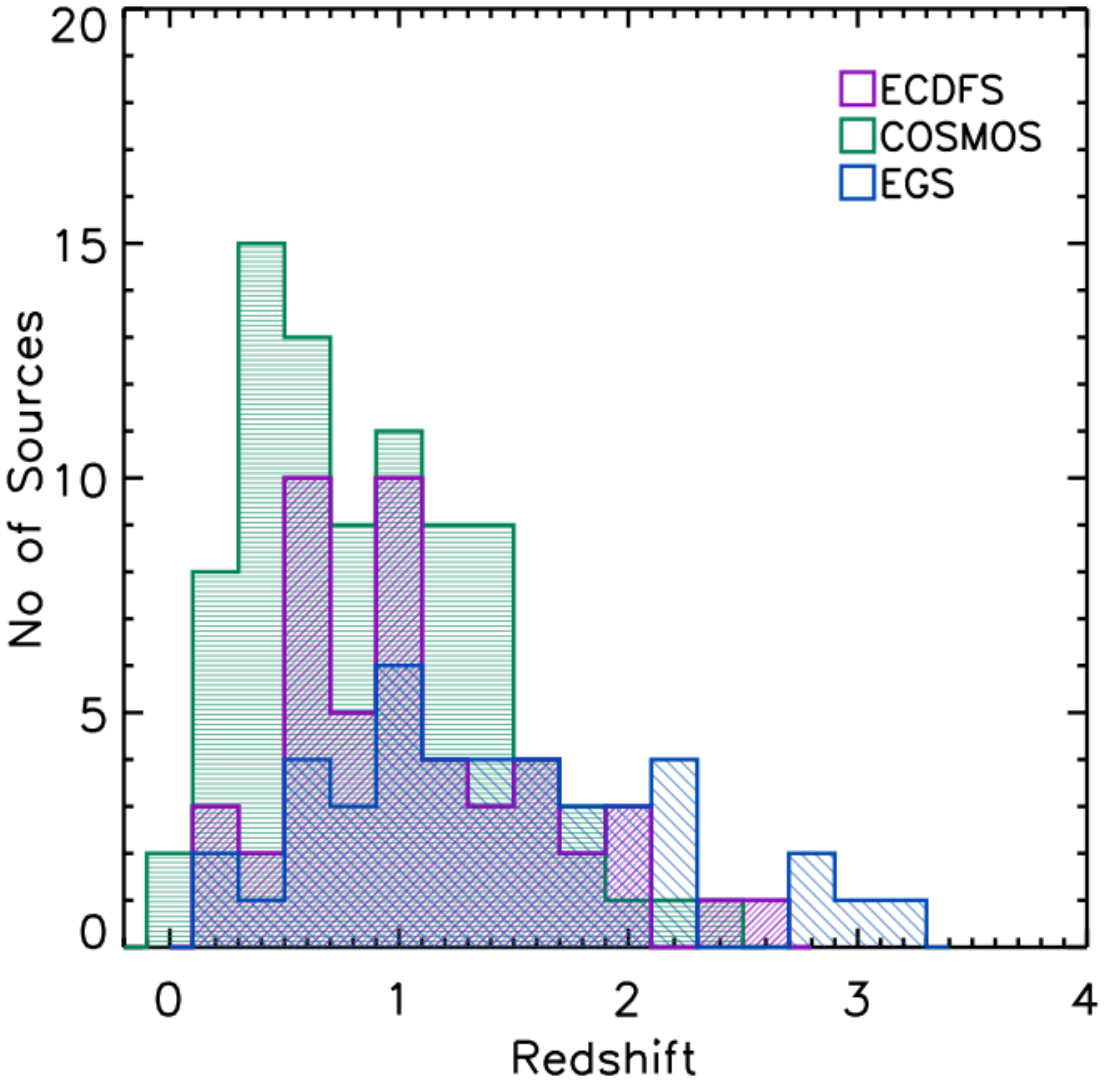}}
%\vspace{-0.5cm}
\caption{Redshift distribution for the sources detected in the various survey fields: E-CDFS (magenta), EGS (blue), COSMOS (green). The redshift distribution ranges between $z=0.044-3.215$, with a mean redshift $<z>\approx1.065$ (median $z\approx1.021$).}\label{fig.z}
\end{figure}

With the low X-ray energy counterpart identification we also obtained the redshift and the optical class associated to those sources from existing catalogues \citep{brusa2010,xue2011,nandra2015}. The redshift distribution of the \nus\ sources detected in the different survey fields is shown is Figure \ref{fig.z}. In the E-CDFS field there are 44 out of 49 redshift identifications for the \nus\ detected sources ($\sim$90\%); of these redshifts 41 are spectroscopic ($\sim$93\%) and three are photometric redshifts ($\sim$7\%; see \citealt{mullaney2015b}). In addition to the redshifts reported in the \nus\ catalogue by \citet{mullaney2015b} we include a redshift identification for NuSTAR J033243-2738.3 (XID 437, in \citealt{lehmer2005}) of $z\approx1.6$, taken from \citet{vignali2015}, which is derived directly from the \xmm\ X-ray spectrum, resulting in a total of 45 redshift identifications ($\sim$92\%). In the COSMOS field 86 out of the 91 \nus\ sources have a \ch\ and/or \xmmn\ counterpart and redshift identification (see \citealt{civano2015}), of which $\sim$93\% are spectroscopic (80/86 sources) and $\sim$7\% are photometric (6/86 sources), yielding the same redshift identification fractions as in the E-CDFS field. In the EGS field all 42 sources have a redshift identification, however the fraction of spectroscopic redshifts is lower than in the other fields: 33 sources have spectroscopic identification ($\sim$79\%), while 9 have photometric redshift ($\sim$21\%). The redshifts of the whole sample span the range of $z=0.044-3.215$, with a mean redshift $<z>\approx1.065$ (median $z\approx1.021$).

From these catalogues we also took the optical classification for our \nus\ sources, where available. The classification of the sources in the E-CDFS and COSMOS fields comes from optical spectroscopy: 18 and 40 sources are classified as BL AGN (FWHM$>$2000~km~s$^{-1}$) and 19 and 29 as NL AGN or emission-line galaxies (included in this paper as ``NL AGN'') in E-CDFS and COSMOS, respectively (see Table \ref{tab.1}). The remainders have no secure optical classification. Given the smaller fraction of optical spectroscopic identifications in the EGS field, we also take into account the classification derived from spectral energy distribution (SED) fitting (\citealt{nandra2015}) and include the sources dominated by unobscured QSO templates (i.e., where QSO emission is $\ge$50\% of the total {in the optical-NIR bands}) as BL AGN (i.e. unobscured, type 1 AGN) and sources dominated by galaxy, or by obscured QSO templates as NL AGN (i.e., obscured, type 2 AGN). We find in total 18 BL AGN and 21 NL AGN (Table \ref{tab.1}). 

\subsection{Spectral extraction}\label{spext}

The \nus\ data have been processed using the standard \nus\ Data Analysis Software (NuSTARDAS, v1.5.1) and calibration files (CALDB version 20131223), distributed within the NASA's HEASARC software (HEAsoft v.6.17\footnote{\url{http://heasarc.nasa.gov/lheasoft/}}). The source spectra have been extracted from each individual pointing using the task \texttt{nuproduct}, from circular extraction regions of 45$\arcsec$ radius {(enclosing $\sim$60\% of the NuSTAR PSF)}; however in crowded regions the extraction radius was reduced (to a minimum of 30$\arcsec${; $\sim$45\% of the PSF}) to minimize the contamination from nearby sources. The background spectra were extracted from four large regions ($\approx150\arcsec$ radius) lying in each of the four quadrants (ccds) in each pointing, removing areas at the position of known bright \ch\ sources ($f_{\rm2-8~keV}>5\times10^{-15}$~\cgs). The source and background spectra, as well as the ancillary files, were then combined using the task \texttt{addascaspec}.   

The \ch\ source spectra were extracted in a consistent way for all the fields, using the 250~ks and 4~Ms \ch\ data in the E-CDFS (\citealt{lehmer2005, xue2016}) and CDFS (\citealt{xue2011}), the 800~ks data in the EGS (\citealt{nandra2015}) field and the 1.8~Ms \ch\ COSMOS (C-COSMOS; \citealt{elvis2009}) and COSMOS-Legacy survey data (\citealt{civano2016}). We used the {\em ACIS Extract} (AE) software package\footnote{The {\em ACIS Extract} software package and Users Guide are available at \url{http://www.astro.psu.edu/xray/acis/acis\_analysis.html}.} \citep{broos2010,broos2012} to extract the source spectra from individual observations using regions enclosing 90\% of the point spread function (PSF); background spectra and relative response matrices and ancillary files were also extracted and then combined by means of the FTOOLS \texttt{addrmf} and \texttt{addarf}. 
 
\section{Data analyses}

\subsection{Spectral analysis}\label{spec}
\begin{figure}
\centerline{
\includegraphics[scale=0.67]{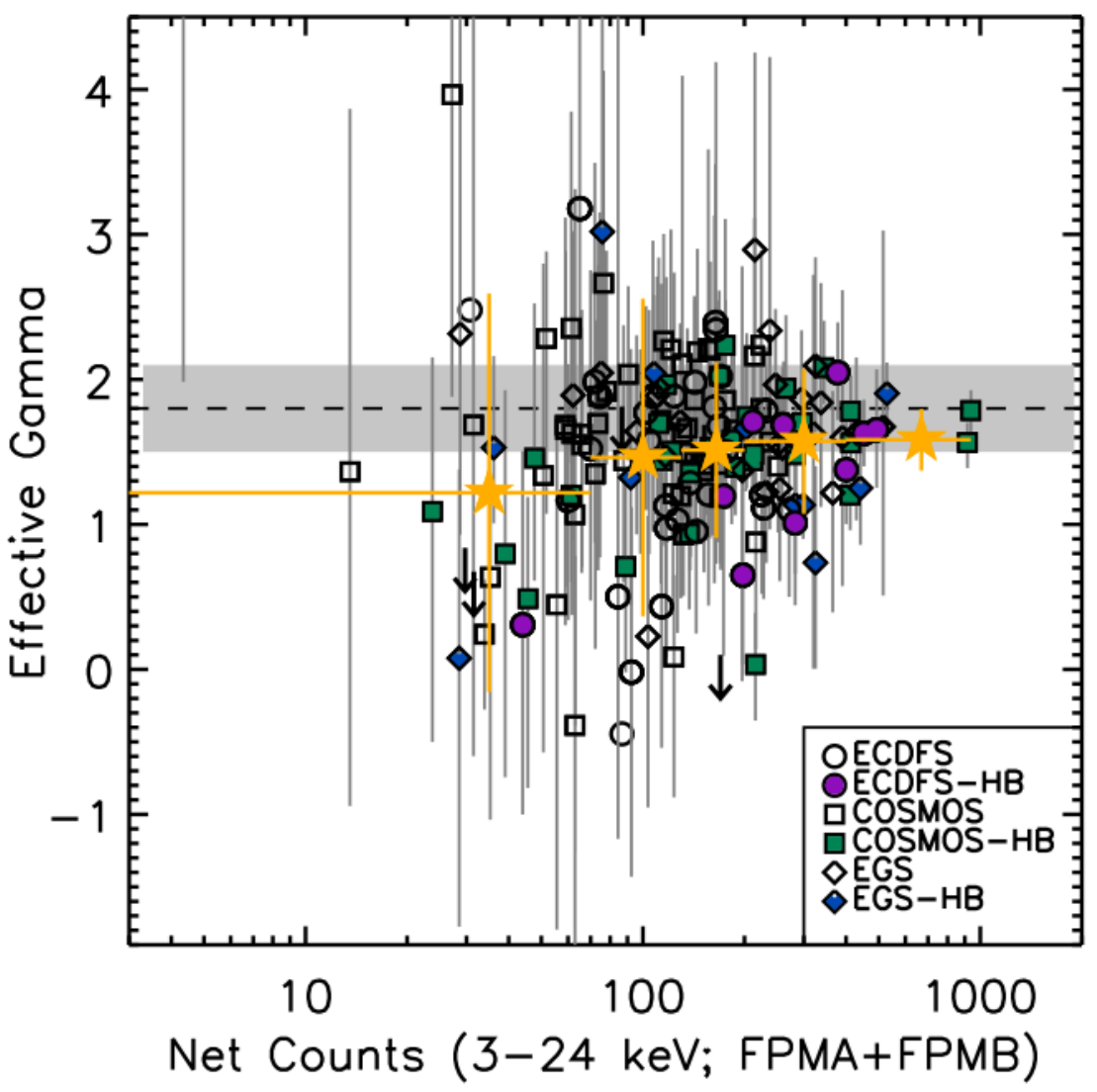}}
\caption{Effective photon index ($\Gamma_{\rm eff}$) versus the net \nus\ counts (FPMA+FPMB) for all the \nus\ detected sources in the E-CDFS (circles), EGS (diamonds) and COSMOS (squares) field. The filled symbols indicate the sources that are significantly detected at $E=8-24$~keV (HB). The shaded region marks the intrinsic photon index $\Gamma=1.8\pm0.3$ typically found for AGN (\citealt{nandra1994,mainieri2002,caccianiga2004, mateos2005a, tozzi2006, burlon2011}). The effective photon index for our sources is derived by fitting a simple power-law model to the \nus\ spectra of each individual source. For many sources the counting statistic is poor and the uncertainties on the $\Gamma_{\rm eff}$ are large. The stars represent the mean $\Gamma_{\rm eff}$ and standard deviation, in different net count bins.}
\label{fig.ctsgamma}
\end{figure}

We first analyzed all the \nus\ spectra ($E\approx3-25$~keV, observed frame) for each individual source to obtain a rough indication of the spectral slope. This is an essential step to produce the composite spectra (see Sect. \ref{stacks}). We therefore fitted all the spectra with a simple power-law model to obtain the effective photon index ($\Gamma_{\rm eff}$; Fig. \ref{fig.ctsgamma}). Since we do not use any redshift information for this initial analysis, we performed the fit for all the \nus\ sources in our sample. The FPMA and FPMB spectra were fitted simultaneously, with a re-normalization factor free to vary, to account for the cross-calibration between the two detectors (\citealt{madsen2015}). Due to the poor counting statistics characterising most of our data (see Fig. \ref{fig.ctsgamma}), the spectra have been lightly binned with a minimum of one count per energy bin and Cash statistic (C-stat; \citealt{cash1979}) was used. The median of the resulting effective photon index is $\Gamma_{\rm eff}=1.57_{-1.26}^{+0.79}$ (the errors correspond to the 5th and 95th percentiles). In Fig. \ref{fig.ctsgamma}, we also show the mean $\Gamma_{\rm eff}$ (and 1$\sigma$ uncertainties) in various net count bins. 

We then performed spectral fitting including the lower energy spectra from \ch. In this case we only include the sources matched to a \ch\ counterpart and with redshift identification (173 sources). The \ch\ data were fitted between 0.5 and 7~keV and an absorbed power-law model including Galactic and intrinsic absorption was used. The Galactic column density was fixed to the mean values of $N_{\rm H}^{\rm Gal}=9.0\times10^{19}$~cm$^{-2}$, $N_{\rm H}^{\rm Gal}=1.05\times10^{20}$~cm$^{-2}$ and $N_{\rm H}^{\rm Gal}=1.79\times10^{20}$~cm$^{-2}$ for E-CDFS, EGS and COSMOS, respectively (\citealt{dickey1990}), while the intrinsic absorption, the photon index ($\Gamma$) and the relative normalization between \ch, FPMA and FPMB spectra were left free to vary. From these spectra we calculated the flux at $3-7$~keV (observed frame), which is the overlapping energy range between the three telescopes, to test whether there is agreement between the \ch\ and \nus\ datasets (since they are not simultaneous), or whether variability might be an issue.  
In general, we found good agreement between the \ch\ and \nus\ fluxes within a factor of two (the mean \nus-\ch\ flux ratio and 1$\sigma$ error is $1.1\pm0.5$), however, at the faint end the \nus\ fluxes are systematically higher than the \ch\ ones. This effect has already been shown by \citet{mullaney2015b} and \citet{civano2015} and is consistent with the Eddington bias, which affects the fluxes close to the \nus\ sensitivity limit.\footnote{For instance, the mean \nus-\ch\ flux ratio is $2.2\pm2.6$ (median $\approx1.4$) for $f_{\rm 3-7~keV}<5\times10^{-14}$~\cgs\ and it increases at the faintest fluxes, due to the effect of the Eddington bias (see Fig. 11 from \citealt{civano2015} and Fig. 5 from \citealt{mullaney2015b}).} Moreover, this effect could be partly due to the \nus\ flux being a blend of multiple sources (see Sect. \ref{data}), while only one \ch\ spectrum was analyzed together with the \nus\ data. For the sources with multiple \ch\ sources within the source matching region, we performed the spectral fit of the \nus\ data with each of the possible \ch\ counterpart and then chose as unique counterpart the \ch\ source with the closest $3-7$~keV flux, which is likely to give the highest contribution to the ``blended'' \nus\ flux.  

\begin{figure}
\centerline{
\includegraphics[scale=0.65]{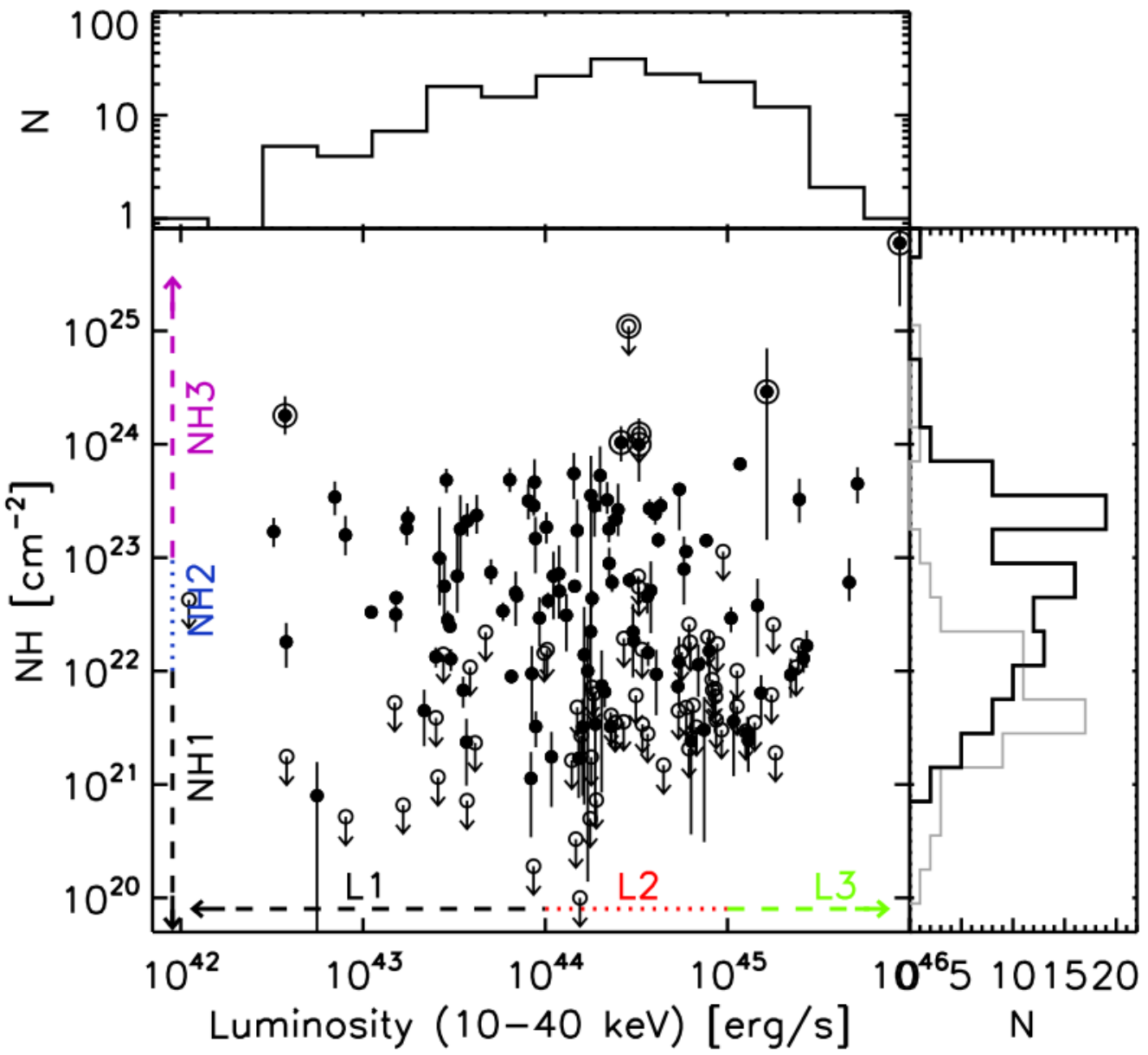}}
\vspace{-0.1cm}
\caption{Hydrogen column density (\nh) versus the 10-40~keV luminosity derived for each individual source using an absorbed power-law model fitted to the \ch\ plus \nus\ spectra, with $\Gamma=1.8$ fixed. The empty symbols indicate the \nh\ upper limits. On the left side of the plot the division in three different \nh\ bins in marked: NH1 (\nh$<10^{22}$~cm$^{-2}$), NH2 (\nh$=10^{22}-10^{23}$~cm$^{-2}$) and NH3 (\nh$>10^{23}$~cm$^{-2}$; see section \ref{nhbins}); at the bottom of the plot the division into three luminosity bins is indicated: L1 ($L_{\rm 10-40~keV}<10^{44}$~\ergs), L2 ($L_{\rm 10-40~keV}=10^{44}-10^{45}$~\ergs) and L3 ($L_{\rm 10-40~keV}\ge10^{45}$~\ergs; see section \ref{lbins}). { Sources that are CT AGN candidates (\nh${\gtrsim10^{24}}$~cm$^{-2}$, within the uncertainties; see Sect. \ref{ct}) are marked with larger open circles.} The top panel shows the distribution of 10-40~keV luminosity of our sample sources. The right panel shows the \nh\ distribution, where the black histogram represents the measured \nh, while the grey histogram represents the \nh\ upper limits.}
\label{fig.nhl}
\end{figure}

\begin{figure*}
\centerline{
\includegraphics[scale=0.44]{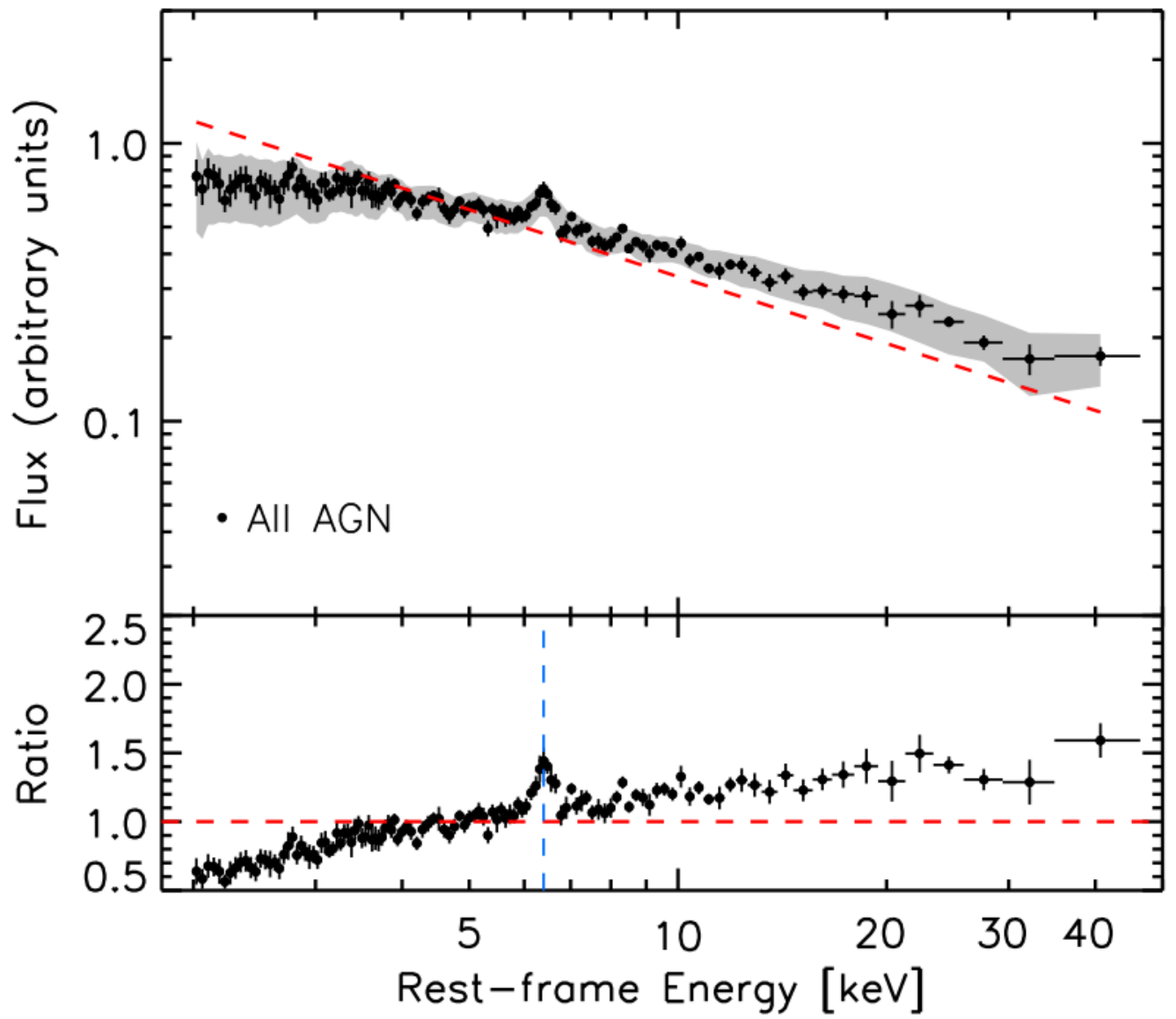}
\hspace{-0.1cm}
\includegraphics[scale=0.44]{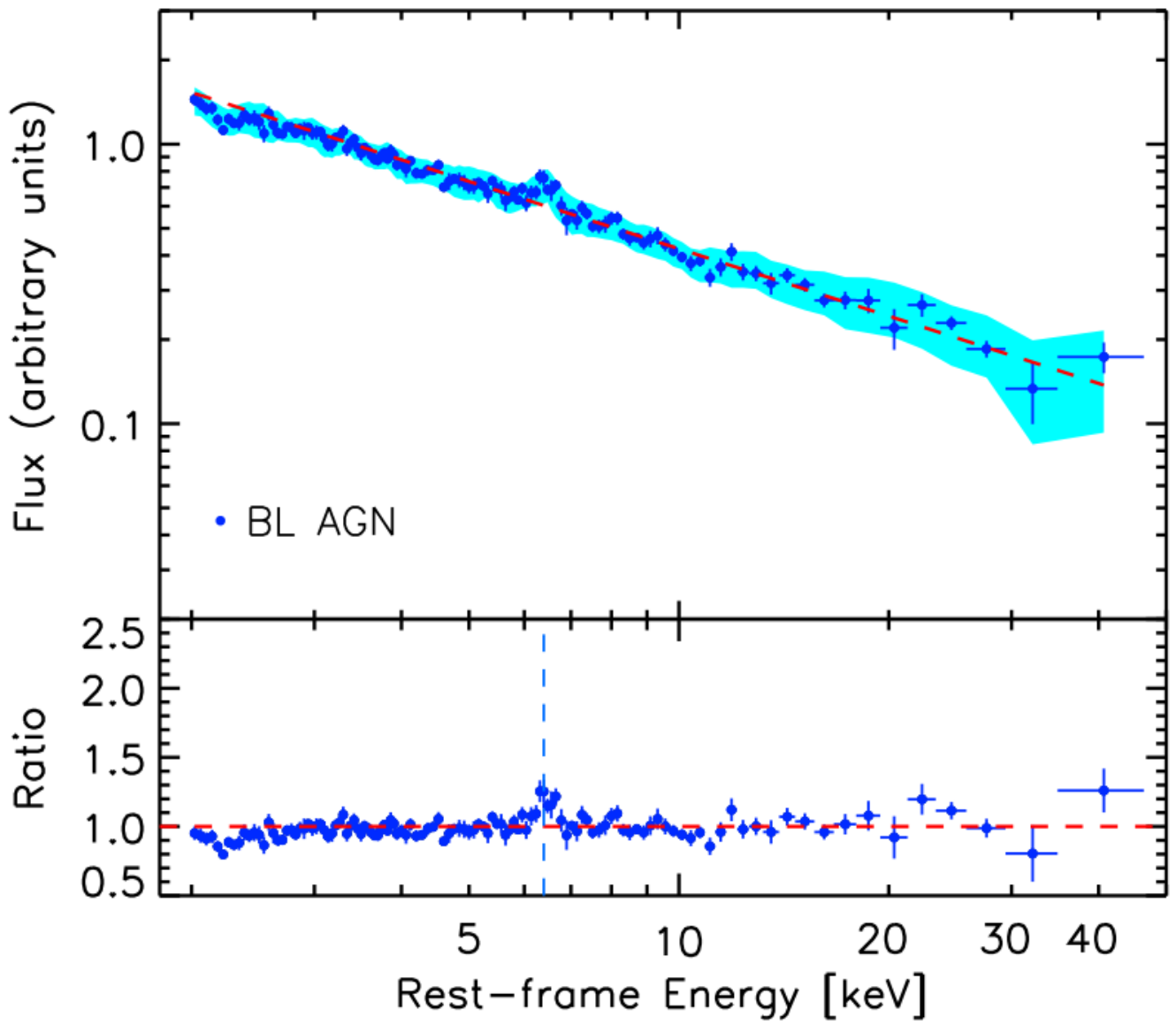}
\hspace{-0.1cm}
\includegraphics[scale=0.44]{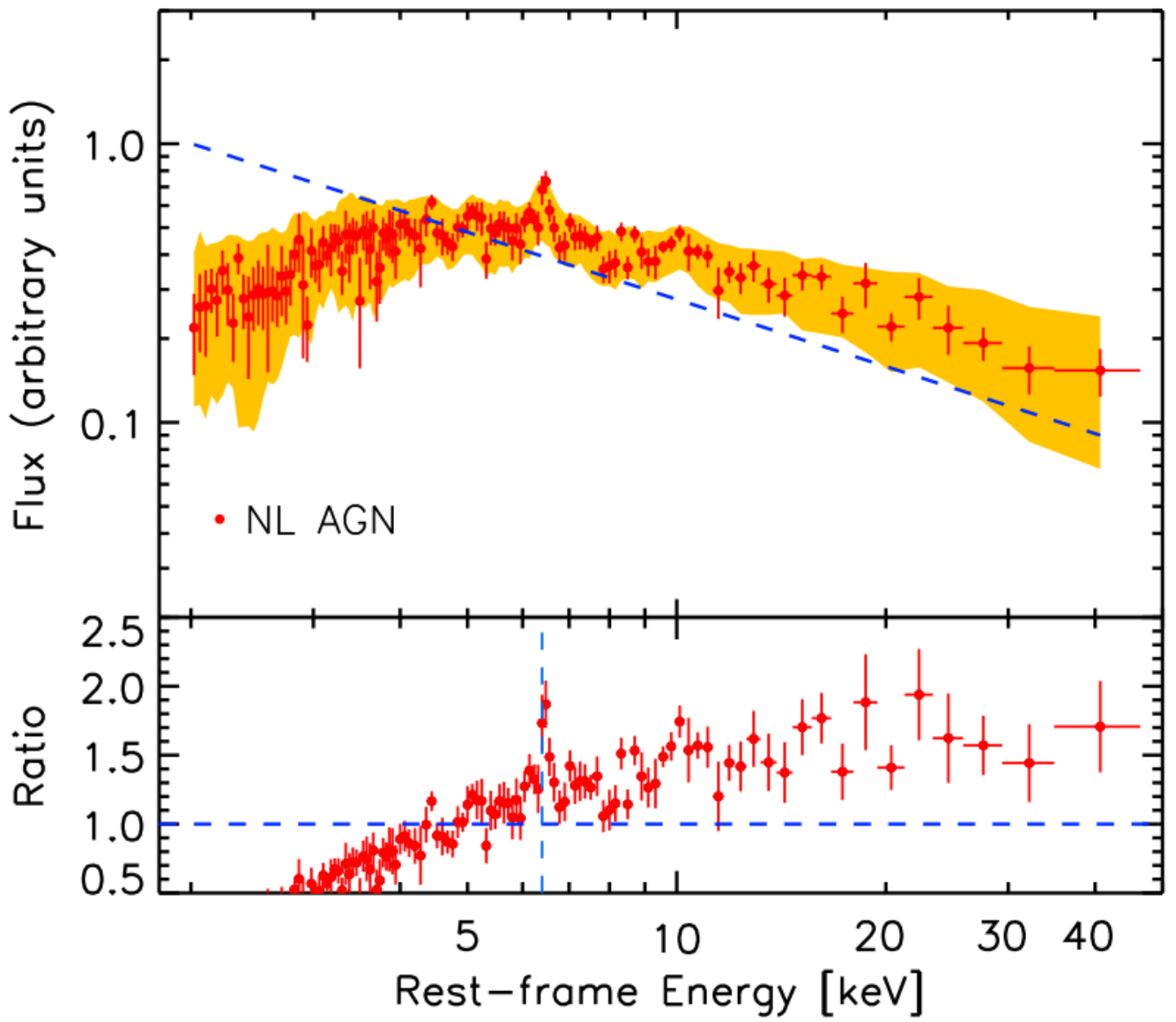}}
%\vspace{-0.1cm}
\caption{Composite spectra in the rest frame $2-40$~keV obtained from the \ch\ and \nus\ data for all the \nus-detected sources in the E-CDFS, EGS and COSMOS fields (black circles; left); the central and right plots show the composite spectra for sources classified in the optical band as BL AGN (blue) and NL AGN (red), respectively. The shaded areas in the three plots represent the range of the composite spectra (1st and 99th percentiles) obtained using a {resampling} analysis and calculating the median spectra 1000 times by randomly selecting subsamples of the sources. The dashed lines in the three plots represent a power law model with a fixed index $\Gamma=1.8$; in the plots the power law is not fitted to the data, but is shown as a reference, by normalising the flux to that of the composite spectra at $E\approx4-5$~keV. The bottom panels show the ratio between the spectra and the power law (dashed line). The vertical dashed lines mark the centroid of the iron K$\alpha$ line at $E\approx6.4$~keV.}\label{fig.mean2}
\end{figure*}

Since in several cases the spectral fit could not provide constraints on both \nh\ and $\Gamma$ simultaneously, we fit the \ch\ and \nus\ data fixing $\Gamma=1.8$, to obtain some constraints on the intrinsic column density. In some cases, when significant residuals are present at soft energies, we added another power-law component to the model, with the spectral slope fixed to that of the primary component (to limit the number of free parameters in the fits; e.g., \citealt{brightman2013,lanzuisi2015,delmoro2016}), but not affected by intrinsic absorption, to account for any soft excess. From these spectra we also calculated the intrinsic X-ray luminosity at $10-40$~keV ($L_{\rm 10-40~keV}$, rest-frame) as we aim to construct composite spectra in different bins of \nh\ and $L_{\rm 10-40~keV}$ (Sects. \ref{nhbins} and \ref{lbins}). In Figure \ref{fig.nhl} we show the \nh\ versus $L_{\rm 10-40~keV}$ distribution for all the analyzed sources. For the composite spectra we will divide the sources in three column-density bins: unabsorbed, \nh$<10^{22}$~cm$^{-2}$ (hereafter ``NH1''), moderately absorbed, \nh$=10^{22}-10^{23}$~cm$^{-2}$ (``NH2'') and heavily absorbed, \nh$>10^{23}$~cm$^{-2}$ (``NH3''; see Section \ref{nhbins}) and in three luminosity bins $L_{\rm 10-40~keV}<10^{44}$~\ergs\ (hereafter ``L1''), $L_{\rm 10-40~keV}=10^{44}-10^{45}$~\ergs\ (``L2'') and $L_{\rm 10-40~keV}\ge10^{45}$~\ergs\ (``L3''; see Section \ref{lbins}).

\subsection{Composite spectra}\label{stacks}

Given the faintness of the sources and the limited counting statistics, in many cases the spectral analysis of the individual sources does not provide constraints on the spectral parameters ($\sim$40\% have \nh\ upper limits; see Fig. \ref{fig.nhl}). We therefore produce composite spectra in order to investigate the average properties of the AGN detected by \nus. 
We use both the \ch\ and \nus\ data together to produce the broad-band composite spectra ($\approx2-40$~keV, rest frame), as the \ch\ data help improving the counting statistics at low energies $E\lesssim7-10$~keV, and allow us to obtain better constraints on the spectral properties than using the \nus\ data alone.  

\subsubsection{Averaging method}\label{ave}

\begin{figure*}
\centerline{
\includegraphics[scale=0.92]{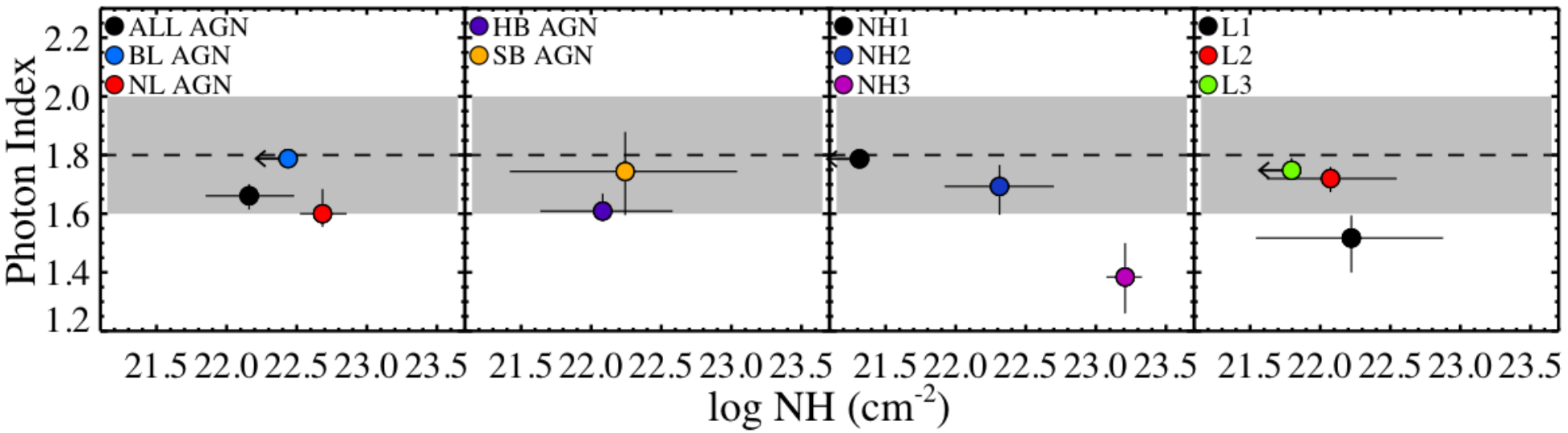}}
\vspace{-0.2cm}
\caption{Photon index vs. hydrogen column density (\nh) derived by fitting the {\tt TORUS} model to each of our composite spectra. From left to right the panels show the results for: i) all the sources (black), BL AGN (blue) and NL AGN (red), as described in Section \ref{blnl}; ii) HB-detected (purple) and SB-detected ({ orange}) AGN (see Section \ref{hbsb}); iii) different \nh\ bins (NH1, black; NH2, blue, and NH3, magenta), described in Section \ref{nhbins}; iv) different $10-40$~keV luminosity bins (L1, black; L2, red, and L3 green), as described in Section \ref{lbins}. The grey shaded area represents the typical AGN photon index of $\Gamma=1.8\pm0.2$.}
\label{fig.gamnh}
\end{figure*}

The composite spectra were produced adopting the averaging method described in \citet{corral2008}. Briefly, using the best-fitting parameters obtained from the spectral fits described in the previous section (\ref{spec}), i.e., an absorbed power law (with both \nh\ and $\Gamma$ free) for the \ch$+$\nus\ spectra, we applied these models to the un-binned, background-subtracted spectra and saved the {\tt unfolded} spectra in XSPEC (v.12.9.0) in physical units (keV~cm$^{-2}$~s$^{-1}$~keV$^{-1}$) in the energy range $3-25$~keV for the \nus\ data and between $E=max(0.5,1.0/(1+z))$ and 7~keV for the \ch\ data. We limit the \ch\ spectra to energies above 1~keV (rest-frame) to minimize the contribution from any soft component, which can create distortions in our composite spectra. Each spectrum was then shifted to the rest frame. To combine the \ch\ and \nus\ data, we renormalize the \ch\ spectra to the \nus\ flux at 3-7~keV first, in order to correct for any flux differences, and apply a correction for the Galactic absorption. We then created a common energy grid for all the spectra, with at least 1200 summed counts per energy bin\footnote{The binning was chosen so that each source contributes, on average, $\gtrsim6-7$ counts per energy bin.}, renormalized all the spectra to the flux in the rest-frame $8-15$~keV energy range, and redistributed the fluxes in each new energy bin using equations (1) and (2) from \citet{corral2008}. The renormalization of the spectra is necessary to avoid the brightest sources dominating the resulting composite spectra. We note that the resulting rescaled spectra preserve their spectral slopes and features. Instead of using the arithmetic mean to calculate the average flux in each new energy bin, as done by \citet{corral2008}, we took the median flux in each bin, as the median is less sensitive to outliers of the flux distribution (e.g., \citealt{falocco2012}). In Appendix \ref{ap} we analyse in details the differences between various averaging methods. To estimate the real scatter of the continuum we performed a {resampling} analysis and produced 1000 composite spectra {drawing random subsamples from the data (e.g., \citealt{politis1994}),} excluding some of the spectra {(at least one)} each time (see Figure \ref{fig.mean2}). For the final composite spectra we took the mean and standard deviation (1$\sigma$) of the distribution of median fluxes obtained from the 1000 composite spectral realizations. In Fig. \ref{fig.mean2} we also show the range of composite spectra (1st and 99th percentiles of the distribution) derived from the {resampling} analysis (shaded areas).   

We note that the unfolding and the averaging process can distort the shape of the spectrum (e.g., \citealt{corral2008,falocco2012}). We therefore performed extensive simulations, which are described in Appendix \ref{ap}, to explore the effects of these distortions on the intrinsic average continuum and therefore to derive reliable results from our analyses.

\section{Results}

In this section we analyse our composite spectra for all of the sources, as well as for sources in different sub-samples. All of the composite spectra have been fit using $\chi^2$ statistic in the energy range $E=3-30$~keV (unless otherwise specified), as $>60$\% of the individual source spectra contribute at this energy range and the spectral simulations (see Appendix \ref{ap}) have shown that some distortions might affect the extremes of the composite spectra ($E\approx2-3$~keV and $E>30$~keV), due to a smaller number of sources contributing to those energy bins. Moreover, the presence of some soft component can contribute to the composites at $E\lesssim3$~keV and therefore cause further distortions in the spectra. For our analysis we adopted three different models: i) an absorbed power law with the addition of a Gaussian emission line at $E\approx6.4$~keV (defined hereafter as ``baseline model''; {\tt wabs$\times$pow$+$gauss}, in XSPEC formalism); ii) a physically motivated torus model, such as {\tt TORUS} (\citealt{brightman2011}), which self-consistently includes the main iron emission lines and Compton scattering { (hereafter ``{\tt TORUS} model'')}; iii) an absorbed power-law model with the addition of a Gaussian emission line and a reflection component ({\tt wabs$\times$pow$+$gauss+pexrav}, in XSPEC formalism; \citealt{magdziarz1995}), in order to constrain the reflection parameter ($R${; hereafter ``{\tt pexrav} model''}).

\subsection{Composite spectra for BL and NL AGN}\label{blnl}

\begin{figure*}
\centerline{
\includegraphics[scale=0.9]{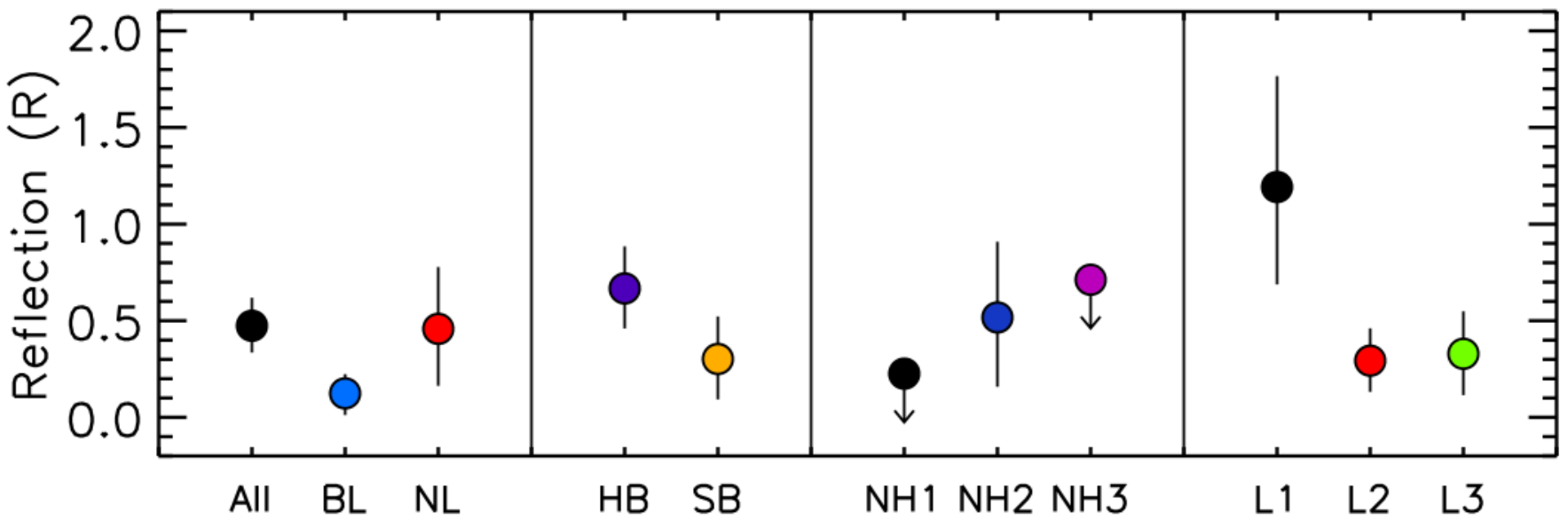}}
\vspace{-0.1cm}
\caption{Reflection fraction (R) derived from the spectral fit of the composite spectra using the baseline model with the addition of a reflection component ({\tt wabs$\times$po$+$pexrav$+$gauss}) and photon index fixed at $\Gamma=1.8$. Symbols are the same as in figure \ref{fig.gamnh}.}
\label{fig.rcl}
\end{figure*}

Figure \ref{fig.mean2} shows the composite spectra from 2~keV to $\sim$40~keV (rest frame) of the \nus$+$\ch\ data. We produced a composite spectrum for all of the 173 sources with redshift identification (spectroscopic or photometric), and also for BL and NL AGN separately. Since the number of sources involved in the BL and NL AGN composite spectra are low compared to the whole sample (see Table \ref{tab.1}), the scatter in the spectra is larger. The \ch\ data significantly improve the counting statistics at $E\lesssim10$~keV over the \nus\ only spectra, and allow us to extend the composites to lower rest-frame energies in order to place constraints on the absorbing column densities, as well as on the intrinsic spectral slope. We initially fit these spectra with our baseline model, with both \nh\ and $\Gamma$ free to vary. The model also includes a Gaussian emission line, as all three composite spectra clearly show an iron K$\alpha$ emission line at $E\approx6.4$~keV. We fixed the line width to $\sigma=0.1$~keV (which is consistent with the values found for an unresolved Gaussian line in stacked spectra; see, e.g., \citealt{iwasawa2012,falocco2013}), while the central energy of the line was left free to vary, to account for possible scatter due to the use of photometric redshifts for some of our sources.\footnote{{We tested the fits also using a narrow Gaussian line fixed at $E=6.4$~keV and width free to vary between $\sigma=0.01-0.2$~keV (e.g., Corral et al. 2008). However, the line width typically pegs at the high limit, as several factors can contribute to broaden the line, such as i) the stacking process (e.g., \citealt{corral2008,falocco2013}); ii) the possible inaccuracy of photometric redshifts; iii) the presence of a broad line component. All these effects could be investigated and disentangled using simulations, however, this is beyond the scopes of this paper. For simplicity, we therefore fix the line width to $\sigma=0.1$~keV and leave the energy free, to account for some of the uncertainties on the line.}} We caution that the hydrogen column densities derived from these spectra do not represent true median values, due to the non-linear nature of the photoelectric absorption.  
The results of our spectral fits are reported in Table \ref{tab.3}. We find a slightly flatter photon index for the composite of all the sources and for the NL AGN ($\Gamma =1.65^{+0.03}_{-0.03}$ and $\Gamma=1.61^{+0.07}_{-0.07}$, respectively) compared to the typical $\Gamma\approx1.8-2.0$. On the other hand, the BL AGN spectral slope is in good agreement with the typical values ($\Gamma=1.78_{-0.02}^{+0.03}$). 

From extensive spectral simulations (see Appendix \ref{ap}), performed to assess the distortions and variations of the true spectral shape, which might occur at different stages of the stacking process, and thus to validate our spectral analysis results, we find that   by simulating unabsorbed \nus\ and \ch\ spectra with a fixed $\Gamma=1.8$, the resulting composite spectrum has a photon index of $\Gamma=1.77-1.83$, in good agreement with the slope of the input simulated spectra. This is true also combining unabsorbed spectra with a range of power-law slopes ($\Gamma=1.6-2.0$; see Appendix \ref{ap}). On the other hand, combining spectra with different levels of X-ray absorption (and therefore different photoelectric cut-off energies), does affect the intrinsic slope of the composite spectra, which becomes slightly flatter ($\Gamma\approx1.72-1.76$; Appendix \ref{ap}.2) than the input $\Gamma=1.8$ of the simulated spectra. This is because the absorption features, which occur at different energies depending on the \nh\ values, are ``smoothed'' during the stacking process, producing an artificial flattening of the spectral slope (as the true intrinsic \nh\ cannot be recovered). We note, however, that in every test we performed with our simulations, we find that this effect is not large enough to explain the relatively flat $\Gamma$ values observed in the composite spectra from the real data. We can therefore assess that the flattening of spectral slope of the NL AGN composite is real and it is significantly different from the slope seen for the BL AGN. An absorbed power-law model with $\Gamma\approx1.61$, in fact, provides a better fit to the data than a slope of $\Gamma=1.8$ at the $>$99.9\% confidence level, according to the $F$-test probability. We then fit the data with the {\tt TORUS} model (\citealt{brightman2011}), where we fixed the torus opening angle\footnote{We chose an opening angle of $\theta_{\rm tor}=30^{\circ}$ because it provides the highest emission-line equivalent width (EW) allowed by this model (\citealt{brightman2011}). Despite this, we find that an extra emission-line component is still needed to reproduce the data.} to $\theta_{\rm tor}=30^{\circ}$ for all of the spectra, while we fixed the inclination angle to $\theta_{\rm inc}=60^{\circ}$ for the BL AGN and $\theta_{\rm inc}=80^{\circ}$ (nearly edge-on) for the NL AGN, as these parameters cannot be constrained in the fit.{We note that changing the $\theta_{\rm inc}$ value for the BL AGN to, e.g., 30$^{\circ}$, makes little difference to the model and to the resulting spectral parameters (the differences are within the parameter uncertainties).} We obtain consistent results with those of the baseline model, with a flattened photon index for the composite spectra of all the sources, and of the NL AGN ($\Gamma=1.66_{-0.05}^{+0.04}$ and $\Gamma=1.60_{-0.05}^{+0.08}$, respectively), compared to that of the BL AGN ($\Gamma=1.79_{-0.01}^{+0.01}$; see Fig. \ref{fig.gamnh}). We note, however, that { significant residuals in the spectra suggest that} the addition of a Gaussian emission line is necessary for all three composite spectra, as the best-fitting {\tt TORUS} model does not fully account for the iron line emission seen in our spectra.  
This suggests that possibly also a Compton-reflection component is not fully represented by this model, which could explain the resulting slightly flat indices.

Indeed, there are two main effects that can produce a flattening in the spectral slope: 1) photoelectric absorption at soft X-ray energies, that can be underestimated and therefore compensated in the fit by a flatter $\Gamma$ (e.g., \citealt{mateos2005a}); 2) Compton reflection contributing to the flux at $E>10$~keV (e.g., \citealt{bianchi2009,ballantyne2011}). The reflection component may arise from either the accretion disc or from cold, dense gas at large distances from the nucleus, such as from the inner part of the putative obscuring torus (e.g., \citealt{ross2005, murphy2009}). %
To verify how much the reflection can be contributing to our composite spectra, we added { then used the {\tt pexrav} model, as} described above. In this model we fixed the photon index of the {\tt pexrav} component to be the same as that of the primary power law $\Gamma=\Gamma_{\rm ref}=1.8$ (assuming that the NL AGN have intrinsically the same spectral slope as the BL AGN), the inclination angle to a mean value of $\rm cos~\theta=0.45$ and the power law cut-off energy $E_{c}=200$~keV (e.g., \citealt{ballantyne2014}). We constrain the reflection parameter\footnote{In our model we force $R$ to be negative, so the {\tt pexrav} component represents a pure reflection component, decoupled from the primary power-law model. However, in the text we report the absolute value of the reflection parameter.} to be $R=0.46_{-0.30}^{+0.32}$ for the NL AGN, while for the BL AGN we obtained $R=0.12_{-0.11}^{+0.10}$ (see Fig. \ref{fig.rcl}). From the composite of all the sources we obtained $R=0.47_{-0.14}^{+0.15}$ ($\Gamma=1.8$ fixed). Although the scatter on the constraints for the NL AGN is relatively large, this shows that a larger contribution from a reflection component in the NL AGN compared to the BL AGN could indeed be responsible for the flattening of the spectral slope. 

From the composite spectra we also derived the equivalent width (EW) of the iron K$\alpha$ emission line at $E\approx6.4$~keV. As stated above, the central energy was left free to vary in the fits to allow for the uncertainties that might arise from using photometric redshifts, while the line width was fixed to $\sigma=0.1$~keV. The resulting centroid of the emission line is always at $E\approx6.4$~keV with typical scatter of $\Delta E\lesssim0.05$~keV. The obtained EWs (reported in Table \ref{tab.3}) from our composite spectra are broadly in agreement with the results from previous works that have investigated the composite X-ray spectra of AGN (e.g., \citealt{corral2008, falocco2012, liu2016}). For the NL AGN we find slightly higher EW values than for the BL AGN, which is consistent with the trend seen for the strength of the reflection component; however, given the uncertainties the trend for the iron line EW is only tentative. Moreover, although some broadening and complexities of the line are visible in the spectra, we only fit a single narrow emission line, since it has been shown that the averaging process and the rest-frame shifting of all the spectra can cause broadening and distortions of the emission line and of the underlying continuum (e.g., \citealt{yaqoob2007, corral2008, falocco2012}). Spectral simulations of the emission line would be required to assess the true nature of these complexities. However, since the detailed properties of the emission line are not the main focus of our paper, we do not attempt more complex analyses.  

\subsection{Comparison between \nus\ hard-band and soft-band detected AGN}\label{hbsb}

\begin{figure}
\centerline{
\includegraphics[scale=0.55]{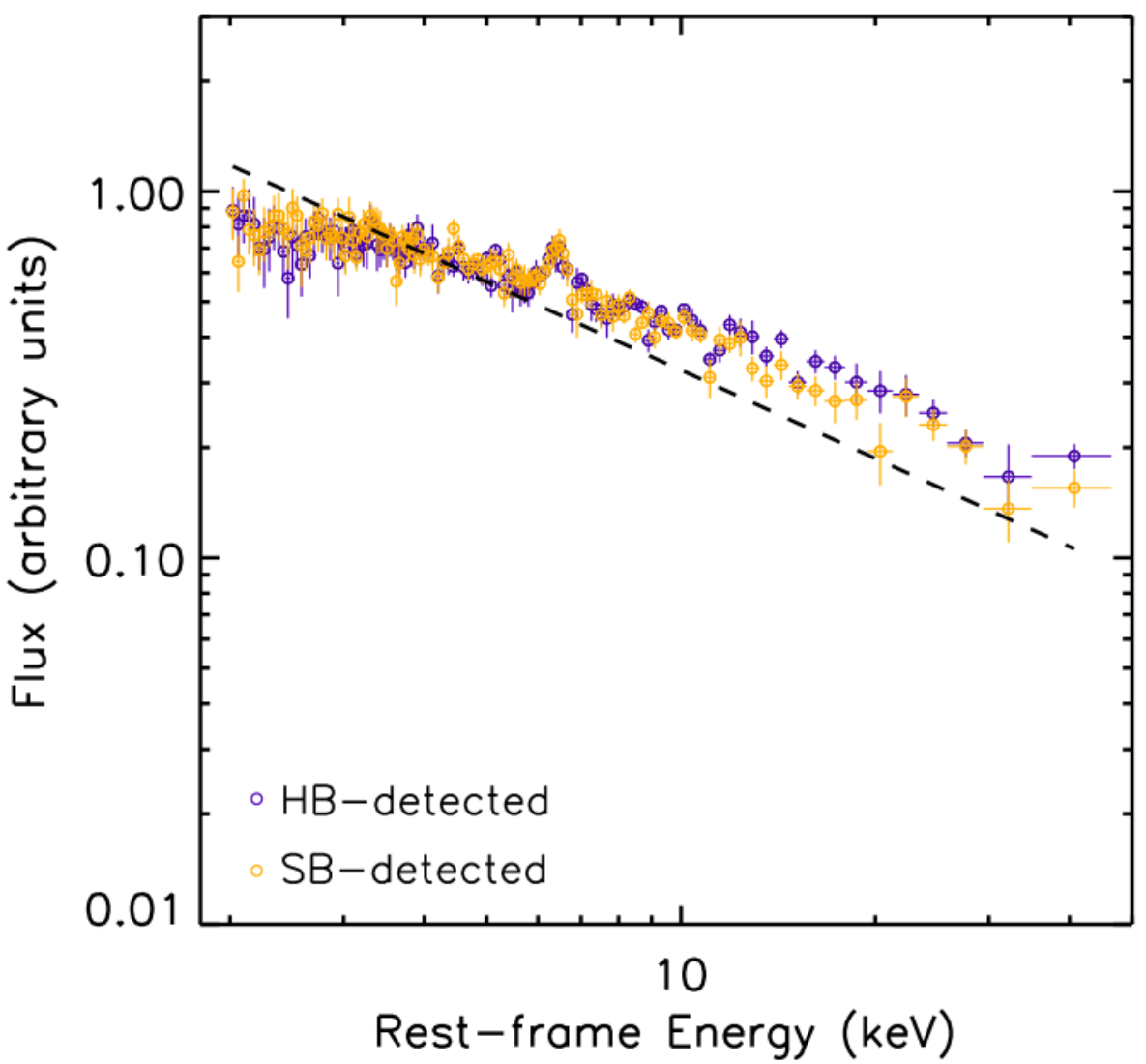}}
%\vspace{-0.2cm}
\caption{Composite spectra in the rest-frame 2-40~keV energy range for sources detected in the \nus\ hard band (HB; purple) and those that are not HB-detected (SB; { orange}). The dashed line in the plot represents a power law with $\Gamma=1.8$.}
\label{fig.hbsb}
\end{figure}
 
With our analyses we also aim to test whether there are significant differences between the intrinsic spectral properties of the sources detected in the \nus\ hard band (HB, $8-24$~keV) and those that formally are not (i.e., they are undetected according to the threshold used by \citealt{mullaney2015b} and  \citealt{civano2015}). We then produced a composite spectrum for all the 64 HB-detected sources (see Table \ref{tab.1}) and for the soft-band (SB; $3-8$~keV) detected sources that are HB undetected (79 sources) and fit the spectra with the models described in the previous section. The spectra are shown in Figure \ref{fig.hbsb}. Using our baseline model, the resulting spectral parameters for the HB composite spectrum are: $\Gamma=1.62_{-0.05}^{+0.05}$ and \nh$=(1.6\pm0.7)\times10^{22}$~cm$^{-2}$, with a fairly weak emission line (EW$=76\pm25$~eV). For the SB composite spectrum the best-fitting parameters are: $\Gamma=1.69_{-0.05}^{+0.05}$ and \nh$=(1.7\pm0.7)\times10^{22}$~cm$^{-2}$, with an EW of the iron K$\alpha$ emission line of EW$=97\pm30$~eV. Adopting the {\tt TORUS} model yields consistent results for the spectral slope and intrinsic \nh\ to our baseline model (see Fig. \ref{fig.gamnh} and table \ref{tab.3}).
\begin{figure*}
\centerline{
\includegraphics[scale=0.7]{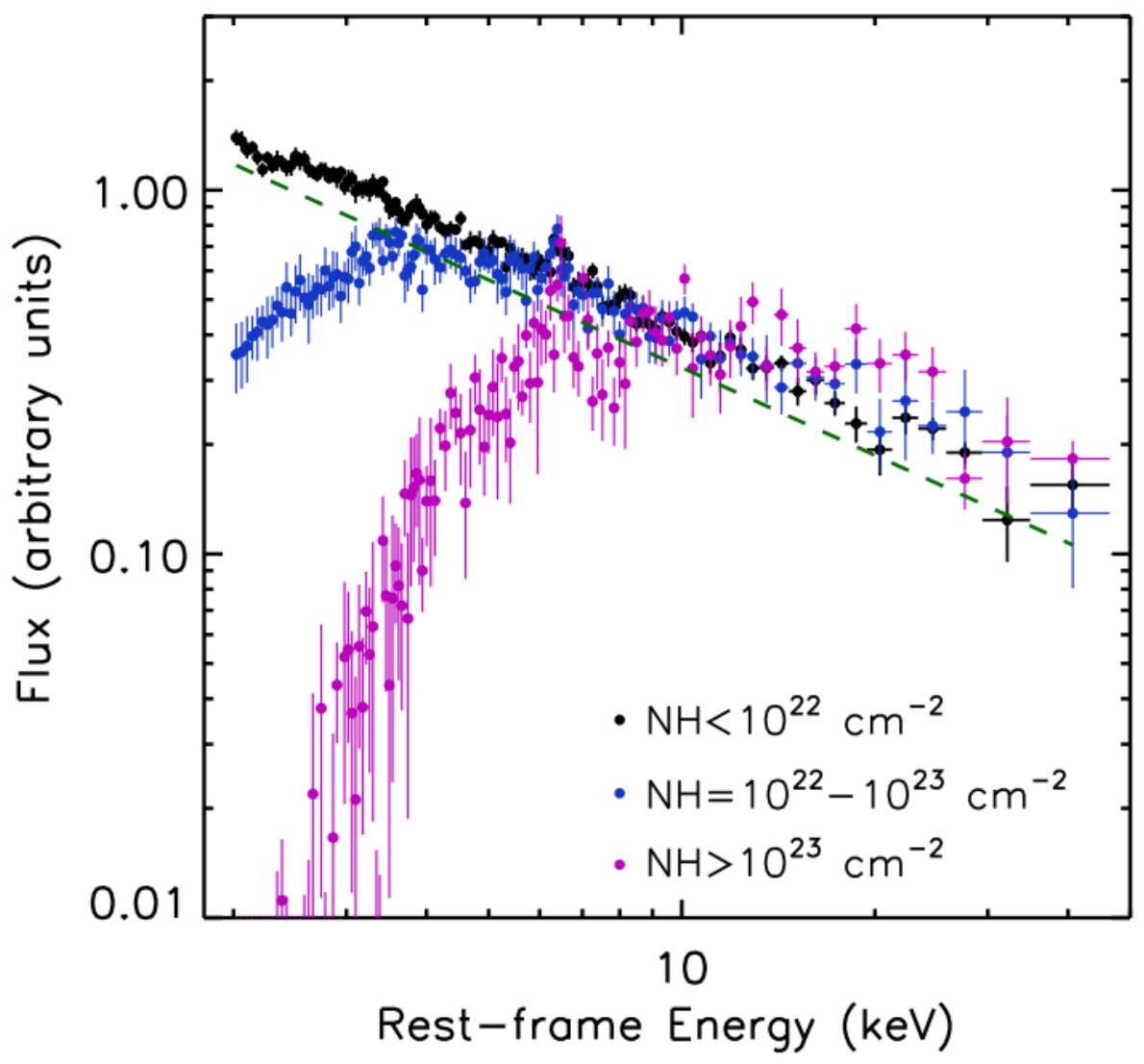}
\includegraphics[scale=0.7]{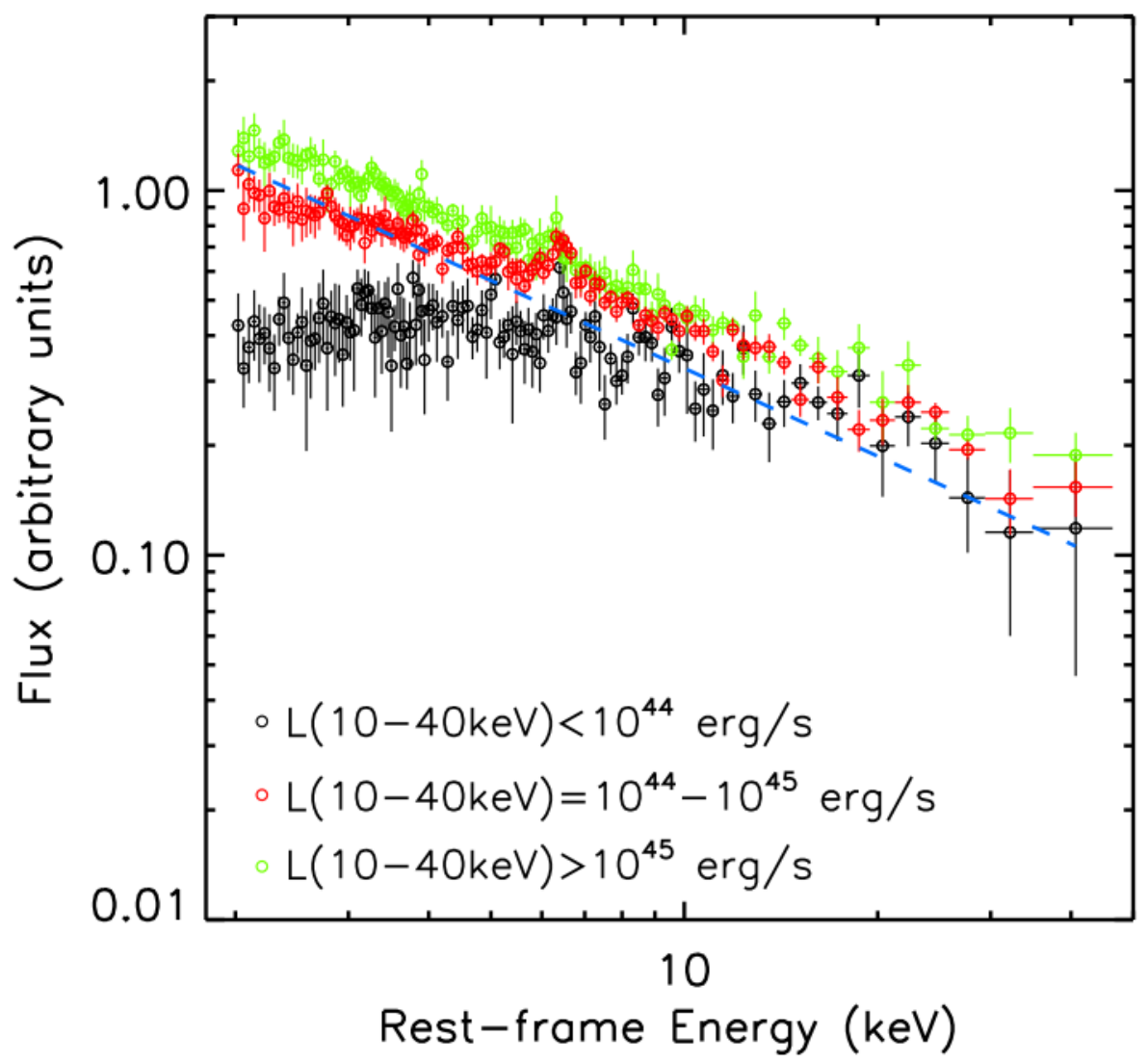}}
\vspace{-0.2cm}
\caption{Left: Composite spectra in the rest-frame 2-40~keV energy range for sources in different \nh\ bins: \nh$<10^{22}$~cm$^{-2}$ (unabsorbed, black), \nh$=10^{22}-10^{23}$~cm$^{-2}$ (moderately absorbed, blue) and \nh$>10^{23}$~cm$^{-2}$ (heavily absorbed, magenta). Right: Composite spectra in the rest-frame 2-40~keV energy range for sources with $L_{\rm 10-40~keV}<10^{44}$~\ergs\ (black), $L_{\rm 10-40~keV}=10^{44}-10^{45}$~\ergs\ (red) and $L_{\rm 10-40~keV}\ge10^{45}$~\ergs\ (green). The dashed line in both plots represents a power law with $\Gamma=1.8$.}
\label{fig.spnh}
\end{figure*}

{ Using the {\tt pexrav} model,} as described in the previous section, we constrain $\Gamma=1.62_{-0.03}^{+0.10}$ and $R<0.39$ ($R=0.67_{-0.21}^{+0.22}$ for $\Gamma=1.8$ fixed) for the HB-detected sources; the spectral parameters are consistent within the uncertainties with those obtained for the SB-detected sources, i.e., $\Gamma=1.69_{-0.03}^{+0.07}$ and $R<0.25$ ($R=0.30_{-0.21}^{+0.22}$ for $\Gamma=1.8$ fixed; Fig. \ref{fig.rcl}). If we attempt to account for the ``artificial'' flattening due to the stacking process, e.g., by fixing $\Gamma=1.76$ (instead of $\Gamma=1.8$; see Appendix \ref{ap}), the constraints on the reflection parameters become $R=0.50_{-0.19}^{+0.20}$ for the HB composite spectrum and $R<0.37$ for the SB composite spectrum. Although there is a hint of an increase of the reflection strength in the composite spectrum of the HB-detected sources, the spectral parameters of the HB- and SB-detected sources are consistent within the uncertainties. These results therefore suggest that the two sub-samples have similar characteristics, with no significant biases toward more obscured or reflection dominated objects in the HB- compared to the SB-detected samples (the fractions of X-ray obscured sources in the two sub-samples is $\sim$52\% and $\sim$46\%, respectively).

\subsection{Average spectral properties for absorbed and unabsorbed AGN}\label{nhbins}

To place better constraints on the contribution from Compton reflection to the average spectra of the \nus\ sources and disentangle its effect from that of absorption in flattening the spectral slope (see Sect. \ref{blnl}), we produce composite spectra in different \nh\ bins. In this way, by knowing \emph{a priori} the median \nh\ of the spectra, we can attribute any hardening of the spectral slope just to the strength of the Compton hump. As an estimate of the intrinsic \nh\ of the sources we used the results from the fitting of the \ch\ and \nus\ spectra for individual sources with $\Gamma=1.8$ fixed (see Fig. \ref{fig.nhl} and Sect. \ref{spec}). We caution that this is a very simple model and provides a crude estimate of the column densities, as more complex models might be needed to fully characterize the individual source spectra (e.g., \citealt{delmoro2014}; Zappacosta et al. 2017, in prep). However such analysis is not feasible for many of the sources given the limited counting statistics. 

We divided the sources in three \nh\ bins (see Sect. \ref{spec}): \nh$<10^{22}$~cm$^{-2}$ (NH1), \nh$=10^{22}-10^{23}$~cm$^{-2}$ (NH2) and \nh$>10^{23}$~cm$^{-2}$ (NH3), including upper limits. To distribute the sources in each bin we approximated the \nh\ and errors of each source to a Gaussian distribution and performed 1000 spectral realizations, randomly picking an \nh\ value from the distribution and assigning the source to one of the three \nh\ bins accordingly. For the sources with a \nh\ upper limit ($\sim$40\% of the sample; see Fig. \ref{fig.nhl}) we assumed a constant probability distribution for the column density values ranging from $log$ \nh$=20.5$~cm$^{-2}$ and the \nh\ upper limit of the source. We excluded two (absorbed) sources due to their particularly strong soft component and/or spectral complexity, which would further increase the scatter of the composite spectra at $E\lesssim4$~keV and around the iron line. The median number of sources in each bin, resulting from the 1000 spectral realizations are: 86 in NH1, 44 in NH2 and 39 in NH3 (see Table \ref{tab.3}). {As described in Sect \ref{ave}, the final average spectra are obtained taking the mean and the 1$\sigma$ standard deviation of the distribution of median fluxes in each energy bin, resulting from the 1000 spectral realizations described above, and {resampling} analysis.} We note that the NH3 bin has typically a smaller number of sources compared to the other two and therefore the scatter in the composite spectrum is larger. The composite spectra in the three different bins are shown in Figure \ref{fig.spnh} (left). 

\begin{table*}
\caption{Spectral parameters of the composite spectra in different \nh\ and luminosity bins.}
\begin{center}
\begin{scriptsize}
\begin{tabular}{l c c c c c c c c c c c c}
\hline
\rule[-1.5mm]{0pt}{3ex}     &                 &\multicolumn{3}{c}{WA$\times$PO+GAUSS}  &  & \multicolumn{2}{c}{TORUS}&  & \multicolumn{3}{c}{WA$\times$PO+GAUSS+PEXRAV} & \\
\cline{3-5}\cline{7-8}\cline{10-12}
\rule[-1.5mm]{0pt}{3ex} Bin & No.  & $\Gamma$       & \nh$^a$    & EW (eV) & $\chi^2/$d.o.f.       & $\Gamma$ & \nh$^a$ & $\chi^2/$d.o.f. & $\Gamma$       & \nh$^a$    & $R$ ($R_{\Gamma=1.8}$) & $\chi^2/$d.o.f.\\ 
\hline
\rule[-1.5mm]{0pt}{4ex}All & 173& $1.65^{+0.03}_{-0.03}$ & 1.8$^{+0.5}_{-0.5}$ & 92$^{+22}_{-22}$ & 86.1/90 & $1.66^{+0.04}_{-0.05}$ & 1.4$^{+0.5}_{-0.4}$ & 88.5/90 & $1.65^{+0.05}_{-0.02}$ & 1.8$^{+0.5}_{-0.5}$ & $<0.17$ ($0.47_{-0.14}^{+0.15}$)  & 86.1/89 \\
\rule[-1.5mm]{0pt}{4ex}BL  & 76 & $1.78^{+0.03}_{-0.02}$ & $<0.3$              & 75$^{+21}_{-21}$ & 94.3/90 & $1.79^{+0.01}_{-0.01}$ & $<2.7$              & 93.7/90 &  $1.84^{+0.08}_{-0.06}$ & $<0.7$     & $<0.64$ ($0.12_{-0.11}^{+0.10}$)      & 91.8/89     \\
\rule[-1.5mm]{0pt}{4ex}NL  & 69 & $1.61^{+0.05}_{-0.05}$ & 5.8$^{+1.0}_{-1.0}$ & 105$^{+41}_{-41}$& 89.5/90 & $1.60^{+0.08}_{-0.05}$ & 4.8$^{+0.8}_{-0.8}$ & 93.0/90 & $1.62^{+0.05}_{-0.07}$ & 5.8$^{+1.0}_{-1.0}$ & $<0.15$ ($0.46_{-0.30}^{+0.32}$) & 89.5/89  \\
\hline
\rule[-1.5mm]{0pt}{4ex}HB  & 64 & $1.62^{+0.05}_{-0.05}$ & 1.6$^{+0.7}_{-0.7}$ & 76$^{+25}_{-25}$ &110.0/90 & $1.61^{+0.06}_{-0.03}$ & 1.2$^{+0.6}_{-0.5}$ &113.1/90 & $1.62^{+0.10}_{-0.03}$ & 1.6$^{+0.8}_{-0.7}$ & $<0.39$ ($0.67_{-0.21}^{+0.22}$) & 110.0/89 \\
\rule[-1.5mm]{0pt}{4ex}SB  & 79 & $1.69^{+0.05}_{-0.05}$ & 1.7$^{+0.7}_{-0.7}$ & 97$^{+30}_{-30}$ & 75.2/90 & $1.74^{+0.13}_{-0.15}$ & 1.8$^{+1.4}_{-1.4}$ & 33.7/90 & $1.69^{+0.07}_{-0.03}$ & 1.7$^{+0.7}_{-0.7}$ & $<0.25$ ($0.30_{-0.21}^{+0.22}$) & 75.2/89 \\
\hline
\rule[-1.5mm]{0pt}{4ex}NH1 & ${86^b}$ & ${1.78_{-0.02}^{+0.02}}$ & ${<0.3}$              & ${62_{-21}^{+21}}$ & { 84.0}/90 & ${1.79_{-0.01}^{+0.02}}$ & $<{0.2}$             & { 84.1}/90 & 1.80${_{-0.03}^{+0.08}}$ & $<{0.5}$              & $<${0.43} ($<${ 0.23}) & { 83.5}/89 \\ 
\rule[-1.5mm]{0pt}{4ex}NH2 & ${44^b}$ & { 1.68}$_{-0.08}^{+0.08}$ & ${2.4}_{-0.9}^{+0.9}$ & ${101_{-43}^{+39}}$ & { 59.2}/90 & 1.69$_{{-0.10}}^{+0.07}$ & ${2.1_{-0.8}^{+0.8}}$& { 58.7}/90 & ${1.72_{-0.09}^{+0.18}}$ & ${2.5_{-1.1}^{+1.2}}$ & $<${1.08} (${0.52_{-0.36}^{+0.39}}$) & { 59.0}/89\\
\rule[-1.5mm]{0pt}{4ex}NH3& ${39^b}$ & ${1.64_{-0.11}^{+0.12}}$& ${28.4_{-3.3}^{+3.6}}$& ${100_{-66}^{+60}}$& { 114.1}/77& ${1.38_{-0.12}^{+0.12}}$ & ${16.2_{-2.6}^{+2.2}}$ & { 184.7/79} & ${1.64_{-0.11}^{+0.14}}$& ${28.3_{-3.3}^{+3.7}}$& $<${ 0.34} ($<${ 0.71}) & { 114.1}/76 \\%3-40
\hline
\rule[-1.5mm]{0pt}{4ex}L1 & ${61^b}$ & ${1.49_{-0.10}^{+0.10}}$ & ${1.8_{-1.3}^{+1.3}}$ & ${115_{-53}^{+53}}$ & { 81.9}/90& ${1.52_{-0.12}^{+0.07}}$ & ${1.7_{-1.1}^{+1.1}}$ & { 81.5}/90 & ${1.51_{-0.06}^{+0.17}}$ & ${1.8}_{-1.3}^{+1.5}$ & $<${ 0.67} (${1.19_{-0.50}^{+0.57}}$)  & { 81.9}/89\\ 
\rule[-1.5mm]{0pt}{4ex}L2 & ${84^b}$ & ${1.71}_{-0.04}^{+0.04}$ & ${1.5_{-0.6}^{+0.6}}$ & ${100_{-17}^{+28}}$  & { 98.8}/90& ${1.72_{-0.05}^{+0.04}}$ & ${1.2_{-0.5}^{+0.6}}$ & { 102.6}/90 & ${1.71_{-0.03}^{+0.10}}$ & ${1.5_{-0.6}^{+0.8}}$ & $<${ 0.34} (${0.29_{-0.16}^{+0.17}}$) & { 98.8}/89\\
\rule[-1.5mm]{0pt}{4ex}L3 & ${22^b}$ & ${1.73_{-0.03}^{+0.05}}$ & $<${ 0.6}              & ${70_{-37}^{+38}}$  & { 77.4}/90 & ${1.75}_{-0.03}^{+0.04}$ & $<{0.6}$              & { 72.0}/90 & ${1.86_{-0.13}^{+0.19}}$ & $<${ 1.5}              & $<${ 1.74} (${0.34_{-0.22}^{+0.22}}$)  & { 70.2}/89 \\
\hline
\hline
\end{tabular}
\end{scriptsize}
\end{center}
NOTES: $^a$ The column density \nh\ is expressed in units of $10^{22}$~cm$^{-2}$; $^b${Median value of the number of sources in each bin, resulting from  1000 spectral realizations with randomized \nh\ or $L_{\rm X}$ values within their error range}.  
\label{tab.3}
\end{table*}%

We again fit the data with our baseline model and with the {\tt TORUS} model, including a Gaussian line. The spectrum in the NH3 bin was fitted between 3.5 and 30~keV, as the spectrum at $E<3.5$~keV has large scatter (see Fig. \ref{fig.spnh}, left), likely due to the presence of a soft scattered component in some of the individual source spectra. The two models yield { fairly} consistent results, which are summarised in Table \ref{tab.3}. From the fit we obtained values of \nh\ consistent with the median values in each bin for all three composite spectra, while we find a significant flattening of the spectral slope for the NH3 sources (with the highest \nh){, especially with the {\tt TORUS} model,} compared to the unabsorbed and moderately absorbed sources (NH1 and NH2), for which the spectral slopes are consistent with the typical values of $\Gamma=1.8\pm0.2$ (see Fig. \ref{fig.gamnh}). We note, however, that the fitting statistic for the NH3 composite spectrum is poor (see table \ref{tab.3}) and therefore the constraints on the spectral parameters for the sources in this \nh\ bin are less reliable than for the sources with lower X-ray absorption.

To constrain the contribution from the Compton-reflection component to the composite spectra, we { used the {\tt pexrav} model, as in the previous sections} (see Sects. \ref{blnl} and \ref{hbsb}). For the NH1 spectrum the best-fit parameters are: ${\Gamma=1.80_{-0.03}^{+0.08}}$ and \nh${<0.5}\times10^{22}$~cm$^{-2}$, with a relative reflection fraction upper limit of ${R<0.43}$ ($\chi^2$/d.o.f.$=${ 83.5}/89; see Table \ref{tab.3}). For the NH2 spectrum the best-fit parameters are: ${\Gamma=1.72_{-0.09}^{+0.18}}$ and \nh$=({2.5_{-1.1}^{+1.2}})\times10^{22}$~cm$^{-2}$, with a relative reflection fraction upper limit of ${R<1.08}$ ($\chi^2$/d.o.f.$=${ 59.0}/89). For the NH3 spectrum the best-fitting solution still favours a slightly flatter photon index of ${\Gamma=1.64_{-0.11}^{+0.14}}$ (although still consistent with typical values within the errors) and \nh$=({28.3_{-3.3}^{+3.7}})\times10^{22}$~cm$^{-2}$, with a low reflection fraction upper limit of ${R<0.34}$ ($\chi^2$/d.o.f.$=${ 114.1}/76). 

\begin{figure}
\centerline{
\includegraphics[scale=0.4]{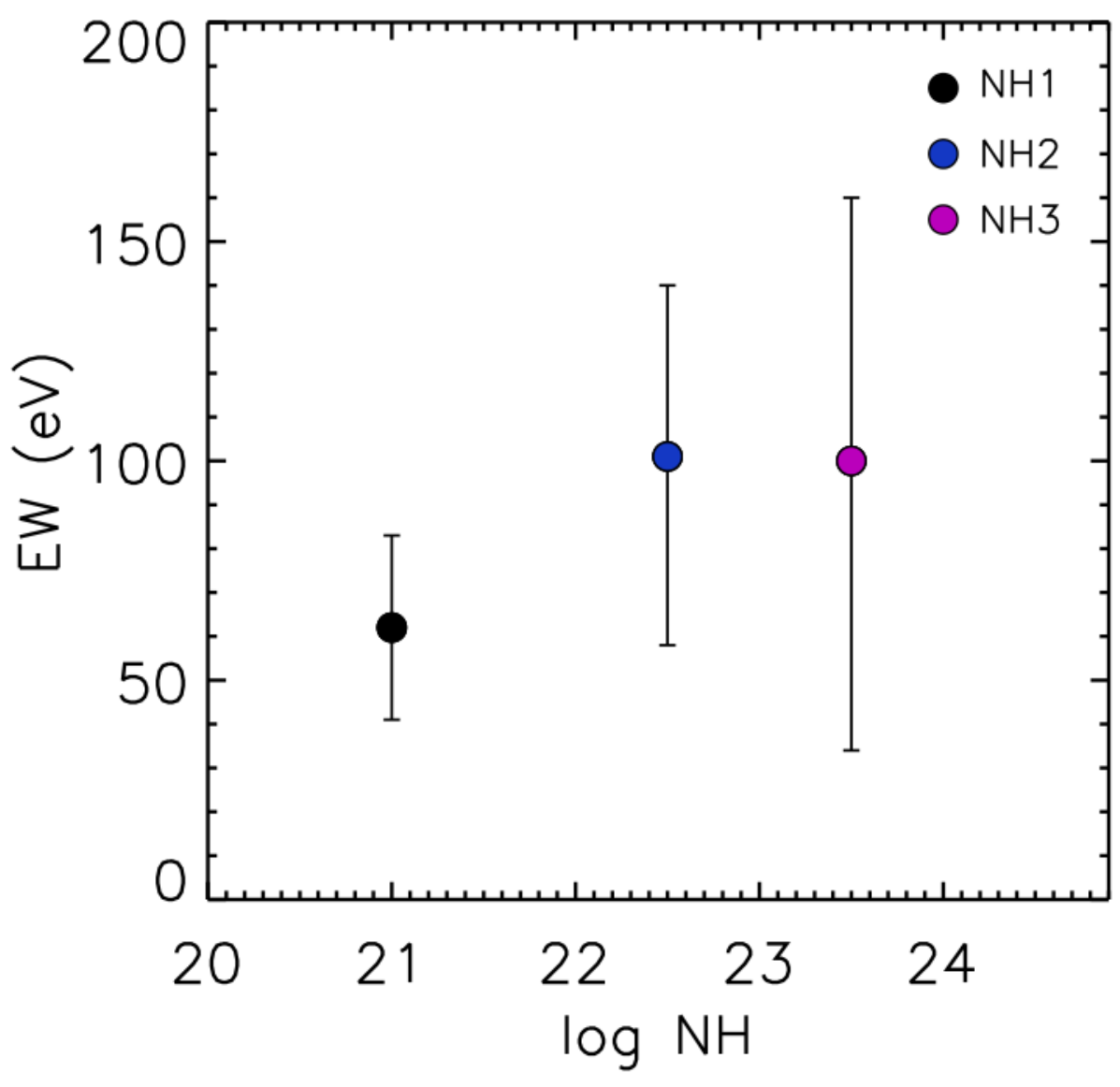}
\hspace{-0.2cm}
\includegraphics[scale=0.4]{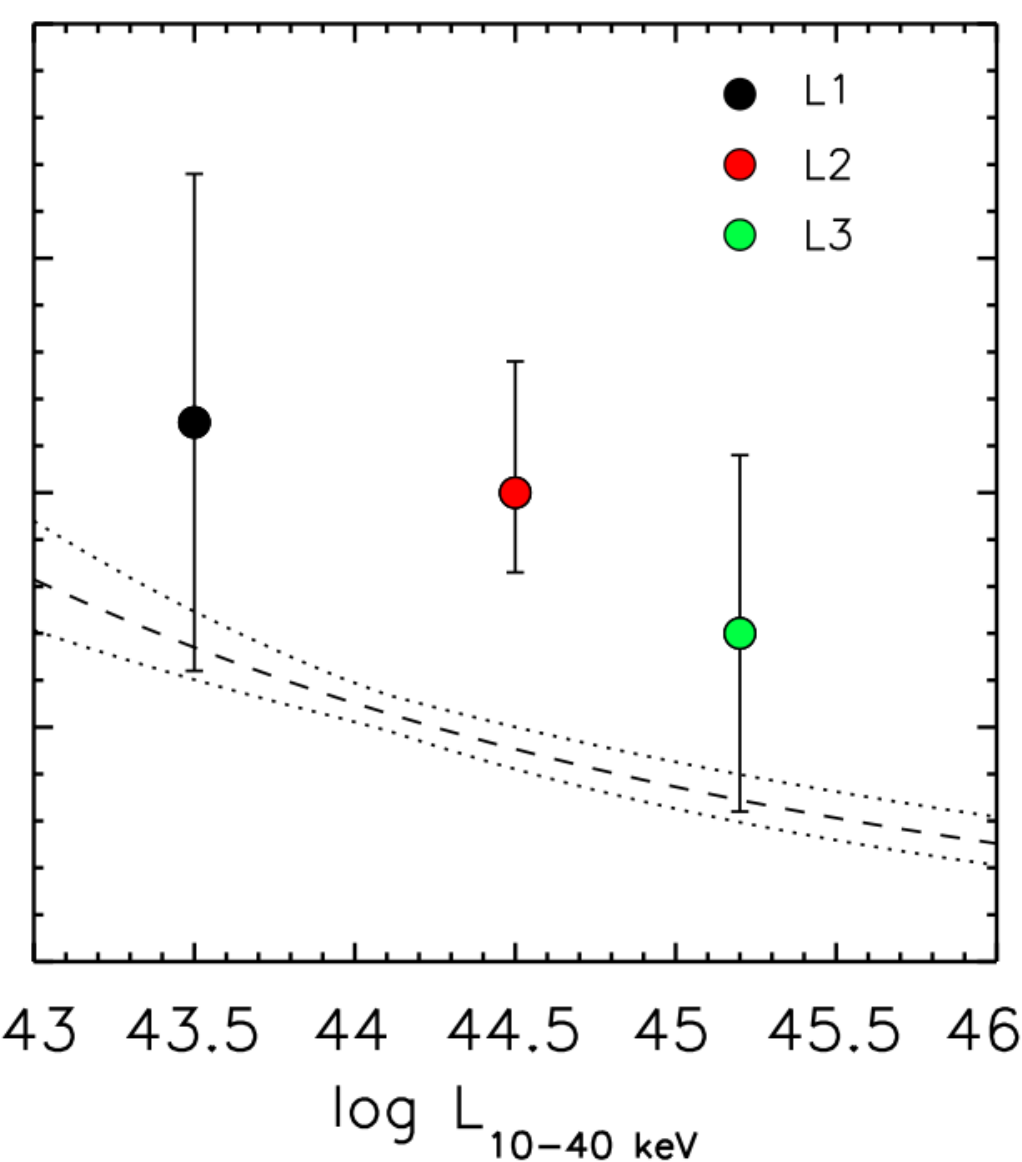}}
%%\vspace{-0.2cm}
\caption{Equivalent width (EW) of the iron K$\alpha$ line measured from our composite spectra. The width of the line was fixed to $\sigma=0.1$ keV in all cases. The left plot shows the EW as a function of \nh, derived from the composite spectra in the NH1, NH2 and NH3 bins. The right panel shows the EW as a function of the 10-40~keV luminosity measured from the composite spectra in the L1, L2 and L3 bins. The dashed and dotted lines show the anti-correlation between the strength of the iron K$\alpha$ line and the X-ray luminosity (``Iwasawa-Taniguchi effect''; \citealt{iwasawa1993}), as measured by \citet{bianchi2007} and the combined errors on the slope and normalization of their best fit. Their $2-10$~keV luminosity has been converted here to the $10-40$~keV luminosity using a power-law model with $\Gamma=1.8$.  
}
\label{fig.ew}
\end{figure}

Since the spectral parameters $\Gamma$ and $R$ are somewhat degenerate when the scatter in the spectra is large (e.g., \citealt{delmoro2014}), we then fixed the photon index to $\Gamma=1.8$, i.e., the intrinsic value found for the unobscured and BL AGN, to obtain better constraints on the reflection fraction. From these spectral fits we obtained ${R<0.23}$ for the NH1 spectrum, ${R=0.52_{-0.36}^{+0.39}}$ for the NH2 spectrum and ${R<0.71}$ for the NH3 spectrum (see Fig. \ref{fig.rcl}; ${R<0.51}$ if we fix $\Gamma=1.75$ to account for the ``artificial'' flattening of the composite spectra with high \nh). The contribution from reflection is typically low in unobscured sources (NH1), while it seems to increase in obscured sources (e.g., \citealt{ricci2011, vasudevan2013}). The relatively poor fit obtained for the NH3 composite spectrum, however, does not allow us to place tight constraints on $R$ and, therefore, to securely assess whether there is a clear dependence between the strength of the Compton reflection and \nh. This is true also for the equivalent width of the iron K$\alpha$ line (see Table \ref{tab.3} and Fig. \ref{fig.ew}, left). Although we would expect and increase of EW with absorption (as the EW is measured against the absorbed continuum), we cannot find a significant dependence, due to the large errors on our EW measurements. We note, however, that the EW of the iron line in the three \nh\ bins has a similar trend to that of the reflection strength (see Fig. \ref{fig.rcl}), despite these two components being fitted independently in our models. This is expected, since the iron line and the Compton hump are two features of the same reflection spectrum.  

\subsection{Luminosity dependence of the X-ray spectral properties}\label{lbins}

To test whether the average spectral parameters of the \nus\ sources change as a function of luminosity, we constructed composite spectra in three different luminosity bins: L1, i.e., sources with $L_{\rm 10-40~keV}<10^{44}$~\ergs, L2 with $L_{\rm 10-40~keV}=10^{44}-10^{45}$~\ergs\ and L3, $L_{\rm 10-40~keV}\ge10^{45}$~\ergs. { Similarly to the method adopted in the previous section, we have approximated the luminosity ${L_{\rm 10-40~keV}}$ and its uncertainties for each source with a Gaussian probability function and performed 1000 realizations randomly picking a ${L_{\rm 10-40~keV}}$ value from the distribution and assigning the source to each luminosity bin. For the luminosity upper limits we assumed again a constant probability function ranging from ${log L_{\rm 10-40~keV}=42.0}$ to the corresponding upper-limit value of the source. From the 1000 realizations the median number of sources in each bin are: 61 in L1, 84 in L2 and 22 in L3, respectively (see Table \ref{tab.3}).} In this case, although the L3 bin has a small number of sources, the composite spectrum has a fairly high S/N as the sources in this bin are brighter and have typically good counting statistics in each individual spectrum. On the other hand, the L1 composite spectrum has relatively high scatter since it comprises the least luminous sources in the sample and therefore the S/N of the individual spectra is typically lower than those in the other luminosity bins. Figure \ref{fig.spnh} (right) shows the composite spectra in the three luminosity bins.

Fitting the spectra with our baseline model, we find that for low-luminosity sources the photon index is flatter (${\Gamma_{\rm L1}=1.49_{-0.10}^{+0.10}}$) than those of higher luminosity sources ($\Gamma_{\rm L2}={1.71}_{-0.04}^{+0.04}$ and $\Gamma_{\rm L3}={1.73_{-0.03}^{+0.05}}$). All the best-fitting spectral parameters are reported in Table \ref{tab.3}.  
Similar results are obtained when using the {\tt TORUS} model to fit the spectra (Fig. \ref{fig.gamnh}). The flattening of the spectrum of the L1 sources might be due to a higher incidence of absorbed sources in this luminosity bin ($\sim$57\%, compared to $\sim$47\% and $\sim$36\% in the L2 and L3 bins, respectively). Indeed, the median value of the column density at $L_{\rm10-40~keV}<10^{44}$~erg~s$^{-1}$ is \nh(L1)$\approx2.5\times10^{22}$~cm$^{-2}$, while for the other two luminosity bins the values are lower: \nh(L2)$\approx8.1\times10^{21}$~cm$^{-2}$ and \nh(L3)$\approx6.4\times10^{21}$~cm$^{-2}$. However, a K-S test on the \nh\ distribution in the three luminosity bins (see Fig. \ref{fig.nhl}) suggests that the distributions are not significantly different ($D(\rm L1-L2)=0.178$ and $Prob(\rm L1-L2)=0.227$ and $D(\rm L1-L3)=0.240$ and $Prob(\rm L1-L3)=0.278$).

Some studies have found a dependence of the photon index with luminosity, where high-luminosity sources show steeper $\Gamma$ than the low-luminosity ones (e.g., \citealt{dai2004, saez2008}). Conversely, other studies have found an anti-correlation between $\Gamma$ and X-ray luminosity (e.g., \citealt{corral2011, scott2011}). Some { of these trends are} probably a consequence of the stronger dependence that has been found for the photon index with Eddington ratio (e.g., \citealt{brightman2013, ricci2013}). The flat photon index we found for the L1 spectrum might be partly due to these correlations. However, the spectral slopes of the sources in the L2 and L3 luminosity bins are pretty much the same (see Fig. \ref{fig.gamnh} and table \ref{tab.3}) and therefore there is no clear dependence of $\Gamma$ with luminosity in our sample (e.g., \citealt{winter2009,brightman2013}). 
\begin{figure*}
\centerline{
\includegraphics[scale=0.6]{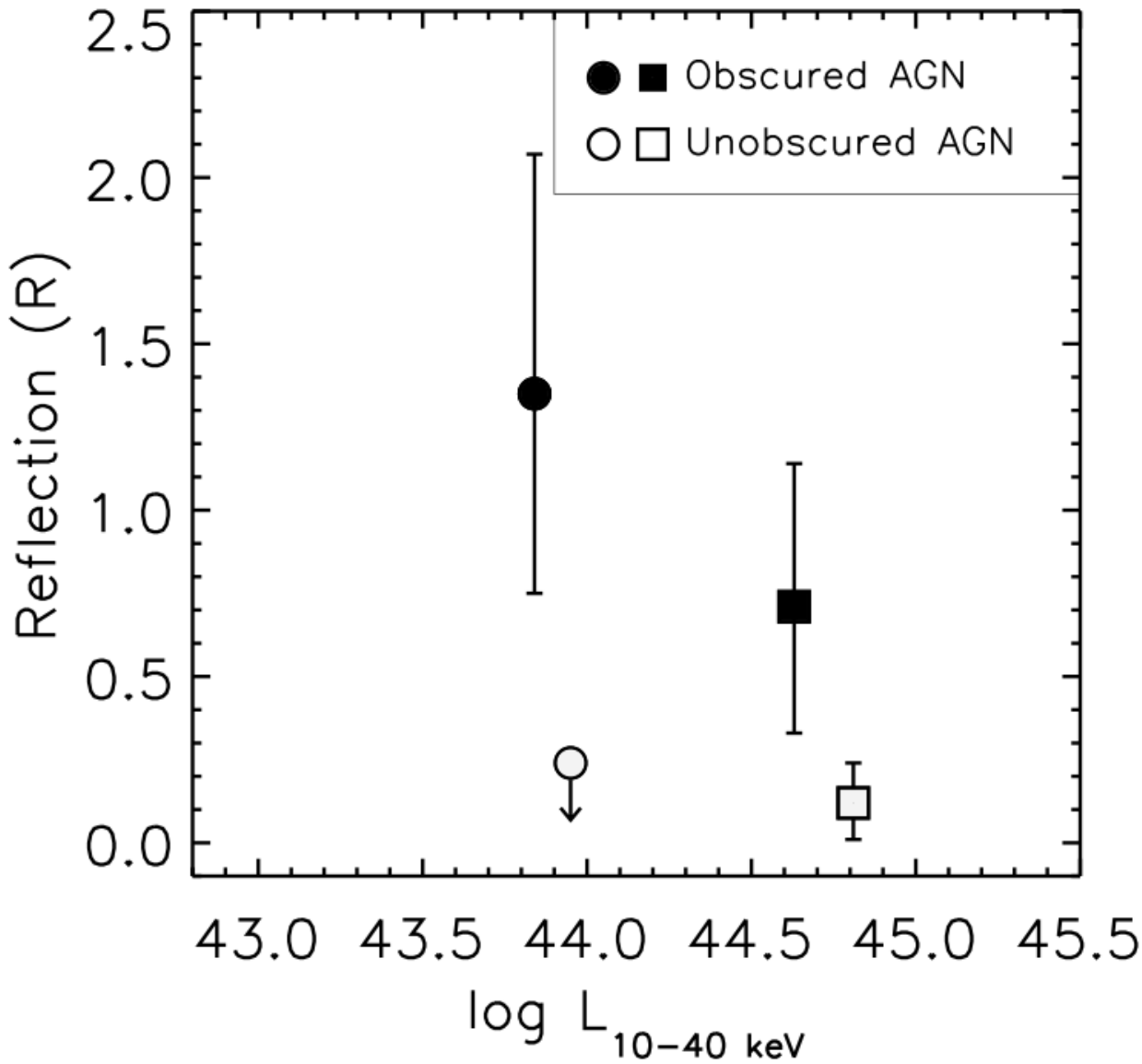}
\includegraphics[scale=0.6]{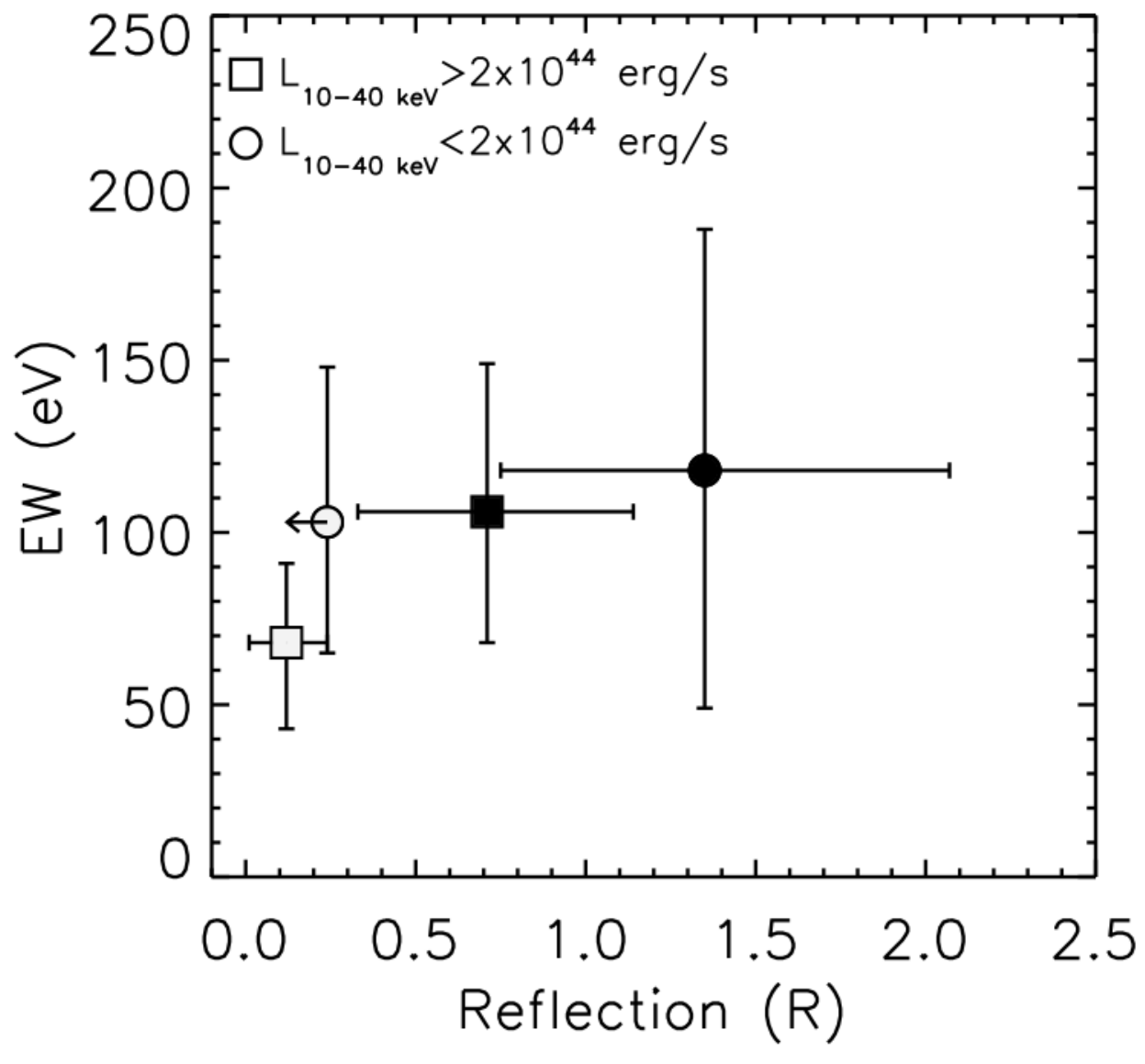}}
%%\vspace{-0.2cm}
\caption{Left: reflection fraction (R) as a function of the rest-frame $10-40$~keV luminosity derived from the composite spectra of the low luminosity sources (${L_{\rm 10-40~keV}<2\times10^{44}}$~\ergs; circles) and high luminosity sources (${L_{\rm 10-40~keV}>2\times10^{44}}$~\ergs; squares), divided into unobscured (i.e., \nh$<10^{22}$~cm$^{-2}$; empty symbols) and obscured (\nh$\ge10^{22}$~cm$^{-2}$; filled symbols) sources. Although the error bars are large, there seems to be a decreasing trend of R both with increasing luminosity and with decreasing \nh. Right: iron K$\alpha$ line equivalent width (EW) vs. the reflection fraction (R), for the same luminosity and \nh\ bins. }
\label{fig.ewR}
\end{figure*}

We therefore investigate whether there is any difference in the amount of Compton reflection contributing to the spectra as a function of luminosity. We fit the {\tt pexrav} model to our spectra, as in the previous sections (Sects. \ref{blnl} and \ref{nhbins}), and we constrain the spectral parameters to be: $\Gamma={1.51_{-0.06}^{+0.17}}$, \nh$=({1.8_{-1.3}^{+1.5}})\times10^{22}$~cm$^{-2}$ and the reflection strength to be ${R<0.67}$ for the L1 sources; $\Gamma={1.71_{-0.03}^{+0.10}}$, \nh$=({1.5_{-0.6}^{+0.8}})\times10^{22}$~cm$^{-2}$ and ${R<0.34}$ for the L2 sources; $\Gamma={1.86_{-0.13}^{+0.19}}$, \nh$< {1.5} \times10^{22}$~cm$^{-2}$ with ${R<1.74}$ for the L3 sources{, which is basically unconstrained}. Fixing the photon index $\Gamma=1.8$ we obtain tighter constraints on the reflection fraction in the three luminosity bins, i.e., ${R=1.19_{-0.50}^{+0.57}}$, ${R=0.29_{-0.16}^{+0.17}}$ and ${R=0.34_{-0.22}^{+0.22}}$ for L1, L2 and L3 sources, respectively (see Fig. \ref{fig.rcl}). While for the low-luminosity sources ($L_{\rm 10-40~keV}<10^{44}$~\ergs) the reflection strength is relatively high, we find that it decreases significantly at luminosities $L_{\rm 10-40~keV}\ge10^{44}$~\ergs\ (e.g., \citealt{bianchi2007, shu2010, liu2016}; however, see also \citealt{vasudevan2013}). Although this trend is only seen fixing the photon index, due to the degeneracies in the spectral parameters, this decrease is significant also if we fit the L1 spectrum with a flatter photon index of $\Gamma=1.75$ to correct for the ``artificial'' flattening discussed in the previous sections, which yields ${R=0.93_{-0.47}^{+0.52}}$. We find hints of a similar, decreasing trend also for the EW of the iron K$\alpha$ line measured from the composite L1, L2 and L3 spectra (see Fig. \ref{fig.ew}, right), although, given the large errors, this trend is not statistically significant. An anti-correlation between the strength of the Fe K$\alpha$ line and the X-ray luminosity has been observed and investigated in several previous studies (e.g., \citealt{iwasawa1993,nandra1997,pagek2004,bianchi2007,ricci2013a}), and it is often referred to as the ``X-ray Baldwin effect'' or the ``Iwasawa-Taniguchi effect'' (IT effect). We compared our results with the anti-correlation found by \citet{bianchi2007} for a sample of radio-quiet, type-1 AGN (dashed and dotted lines in Fig. \ref{fig.ew}, right). We converted their $2-10$~keV luminosity into the $10-40$~keV luminosity assuming a power-law model with $\Gamma=1.8$. The slope of the anti-correlation is consistent with our results, however, the EWs we find in our spectra are typically larger compared to those found by \citet{bianchi2007}. This is likely due to the different types of analyses and different type of sources (as we also include absorbed AGN) used in the two papers, as the measurements from our stacking analyses might be biased toward higher values of EW compared to the distribution found from individual sources. Moreover, the emission line in our composite spectra is broadened due to the rest-frame shifting of the spectra (see Sect. \ref{ave}){, thus likely  increasing} the measured EW. 
\begin{table*}
\caption{Spectral parameters of the composite \ch$+$\nus\ spectra for the low and high-luminosity sources, further divided into unobscured (unob) and obscured (obs), fitted with our \tt{pexrav} model: {\tt wabs$\times$pow+gauss+pexrav}.}
\begin{center}
\begin{tabular}{c c c c c c c c c}
\hline
\rule[-1.5mm]{0pt}{4ex} Bin & No.$^{a}$ & log $L_{\rm10-40~keV}^{b}$ & $z^{c}$ & $\Gamma$ & \nh$^{d}$ & EW (eV) & $R$ ($R_{\rm\Gamma=1.8}$) & $\chi^2/d.o.f.$\\
\hline
\rule[-1.5mm]{0pt}{4ex}low-L unob & 39 & 43.95 & 0.698 & ${1.80_{-0.07}^{+0.10}}$ & ${<1.4}$ & ${103_{-38}^{+45}}$ & ${<0.36}$ (${<0.24}$) & 71.2/89 \\ 
\rule[-1.5mm]{0pt}{4ex}low-L obs & 48 & 43.84 & 0.660 &${1.42_{-0.10}^{+0.12}}$ & ${6.5_{-1.9}^{+1.2}}$ & ${118_{-69}^{+70}}$ & ${<0.30}$ (${1.35_{-0.60}^{+0.72}}$) & 91.8/89 \\
\rule[-1.5mm]{0pt}{4ex}high-L unob & 46 & 44.81 & 1.342 & ${1.85_{-0.05}^{+0.06}}$& ${<0.5}$ & ${68_{-25}^{+23}}$ & ${0.39_{-0.31}^{+0.40}}$ (${0.12_{-0.11}^{+0.12}}$) & 83.4/89\\
\rule[-1.5mm]{0pt}{4ex}high-L obs & 33 & 44.63 & 1.375 & ${1.73_{-0.17}^{+0.22}}$ & ${8.4_{-2.2}^{+2.3}}$ & ${106_{-50}^{+62}}$ & ${<1.65}$ (${0.71_{-0.38}^{+0.43}}$) & 42.5/89\\
\hline
\hline
\end{tabular}
\end{center}
Notes: $^a$ Number of spectra used for the composite{; this is the median value obtained from the 1000 spectral realizations}; $^b$ Logarithm of the median X-ray luminosity ($10-40$~keV) of the sources in each bin; $^{c}$ Median redshift of the sources in each bin; $^d$ The hydrogen column density is expressed in units of $10^{22}$~cm$^{-2}$. 
\label{tab.4}
\end{table*}%

If the IT effect is due to a decrease of the covering factor with increasing AGN luminosity (e.g., \citealt{bianchi2007}), this could also explain the drop of the reflection strength, as in high-luminosity sources there is less material close to the nucleus obscuring/reflecting the intrinsic X-ray emission, thus producing a weaker Compton reflection spectrum.
We caution, however, that we cannot exclude that the drop of $R$ at high luminosities could be partly due to an evolution of the spectral properties with redshift, as the mean redshifts of the sources in the L1, L2 and L3 luminosity bins are $ \langle z\rangle=0.55$, $ \langle z\rangle=1.17$ and $ \langle z\rangle=1.86$, respectively.

\subsection{Compton reflection strength and iron K$\alpha$ line}\label{Rline}

Even dividing the sources in bins of column density, or X-ray luminosity, it is not possible to fully understand how the average spectral properties, such as the strength of the iron K$\alpha$ line and the Compton reflection, depend on these two quantities, as \nh\ and $L_{\rm X}$ are somewhat linked. For instance, the NH1 bin contains a larger fraction of high-luminosity sources ($\sim$74\%) compared to the NH2 and NH3 bins ($\sim$67\% and $\sim$59\%, respectively). Similarly, the L1 bin has a larger fraction of absorbed sources than the high-luminosity bins (see Sect. \ref{lbins}). To further investigate how the Compton reflection might depend on luminosity and/or \nh\ independently, we divided the sources into two luminosity bins, ${L_{\rm10-40~keV}<2\times10^{44}}$~erg~s$^{-1}$ and ${L_{\rm10-40~keV}\ge2\times10^{44}}$~erg~s$^{-1}${; the separation was chosen to have a comparable number of sources within the low and high luminosity bins (see table \ref{tab.4}). Within} these bins we produce separate composite spectra for the unabsorbed (\nh$<10^{22}$~cm$^{-2}$) and absorbed (\nh$\ge10^{22}$~cm$^{-2}$) sources. We adopted the same randomization method described in sections \ref{nhbins} and \ref{lbins} to assign sources to each bin. Since our aim here is to investigate the reflection component, we only fit the composite spectra with the {\tt pexrav} model. The resulting parameters are reported in table \ref{tab.4}. In Figure \ref{fig.ewR} (left) we show the reflection parameter (R) derived from this analysis as a function of the rest-frame $10-40$~keV luminosity; for each composite spectrum, we plot the median $L_{\rm 10-40~keV}$ of the sources. The uncertainties on the derived spectral parameters are quite large, especially for the obscured sources, and therefore it is not possible to derive a statistically significant dependence of $R$ with luminosity or \nh; however, there is an indication of a decreasing trend with luminosity.  
Moreover, the unobscured sources seem to have typically a smaller reflection fraction than the obscured ones. These trends seem to be confirmed by the strength of the iron line measured from these spectra, as shown in Figure \ref{fig.ewR} (right), where the iron line EW is plotted against $R$. The strength of the emission line for obscured and unobscured sources, seems to follow a similar behaviour as the Compton reflection strength. Indeed, a Spearman Rank Correlation Test, performed on 10000 simulated datasets accounting for the uncertainties of the data points, indicates that there is a strong correlation, as the resulting correlation coefficient is ${\rho_{\rm s}=0.80}$; however, given the small number of data points and the large uncertainties, the null hypothesis probability is ${P=0.38}$. 

\section{Compton-thick AGN in the \nus\ survey}\label{ct}

From the spectral analysis of the \ch\ and \nus\ data of the individual sources in our sample using a simple absorbed power-law model with $\Gamma=1.8$ fixed (Sect. \ref{spec}), we identified seven sources with column densities consistent (within the errors) with CT values (\nh$\gtrsim 10^{24}$~cm$^{-2}$, see Fig \ref{fig.nhl}). One of these sources (\nus\ ID 330 in \citealt{civano2015}) was indeed identified as a CT AGN by \citet{civano2015} from the spectral analysis of the \nus, \ch\ and \xmmn\ data available for this source. For the other six sources (\nus\ ID 8 in E-CDFS, IDs 129, 153, 189 and 216 in COSMOS and ID 22 in EGS), which have typically lower S/N spectra both with \nus\ ($<$150 net counts, FPMA+FPMB) and \ch\ ($<$100 net counts), we cannot place tight constraints on the spectral parameters (for two of these sources we only have an upper limit on \nh), and thus we are not able to confirm their CT nature, individually. The sources \nus\ IDs 129, 153, 189 and 22, which are matched to the \ch\ counterparts (CIDs) 284, 1021, 875 in COSMOS (\citealt{elvis2009}) and CID 718 in EGS (\citealt{nandra2015}), respectively, have been already identified as heavily obscured sources by \citet{brightman2014}, with only CID 718 having a column density consistent with CT values within the uncertainties. 

Performing the spectral analysis of the individual sources, with the addition of \xmmn\ data to the \nus\ and \ch\ data, Zappacosta et al. (2017, in prep) constrained IDs 129 and 216 in COSMOS to be heavily obscured, but not CT AGN. We therefore exclude these sources from our list of CT AGN candidates. 

For the remaining five sources, we therefore performed joint-spectral fitting (see, e.g., \citealt{alexander2013}), in the attempt to place better constraints on the average spectral parameters, as well as the reflection fraction, of these CT AGN candidates. In this case we do not produce a composite spectrum because, given the small number of sources, it would be dominated by the uncertainties. We note, however, that the joint-spectral fitting method is analogous to the spectral fitting of the composite spectra for deriving the average spectral parameters. We initially fit the \ch\ ($E=0.5-7$~keV) and \nus\ data ($E=3-25$~keV) using an absorbed power-law model including Galactic and intrinsic absorption, with $\Gamma$ and \nh\ free to vary, and a {narrow} Gaussian line with $E=6.4$~keV and $\sigma=0.05$~keV, analogous to the baseline model we used to fit the composite spectra. Since we include softer energies compared to those of the composite spectra, we also include a soft component, parameterised, for simplicity, as a power law with the same photon index as the primary, absorbed power law: {\tt wabs$\times$(po$+$zwabs$\times$po$+$zgauss)}. With this model, the resulting best-fitting parameters are: $\Gamma=1.81_{-0.35}^{+0.38}$ and \nh$={(1.5_{-0.6}^{+0.8})\times10^{24}}$~cm$^{-2}$,  
i.e., consistent with the CT regime, as derived originally from the spectral fit of the individual sources. Due to the low counting statistics of the spectra, the iron emission line could not be constrained, as its equivalent width resulted in a limit of EW${<0.7}$~keV.  
This is not necessarily in contrast with the presence of CT absorption, as some CT AGN have been reported to have unusually weak Fe lines (e.g., \citealt{gandi2017}). To allow for the spectral complexity expected for these heavily-obscured sources, we used the {\tt TORUS} model, with an inclination angle fixed at $\theta_{\rm inc}=80^{\circ}$ and a torus opening angle fixed at $\theta_{\rm tor}=60^{\circ}$, with the addition of a soft scattered power law. The resulting photon index and hydrogen column density, $\Gamma={1.91_{-0.41}^{+0.59}}$ and \nh$=({9.6_{-4.9}^{+8.6}})\times10^{23}$~cm$^{-2}$, are in good agreement with the results from the previous model, within the uncertainties. 
\begin{figure*}
\centerline{
\includegraphics[scale=0.67]{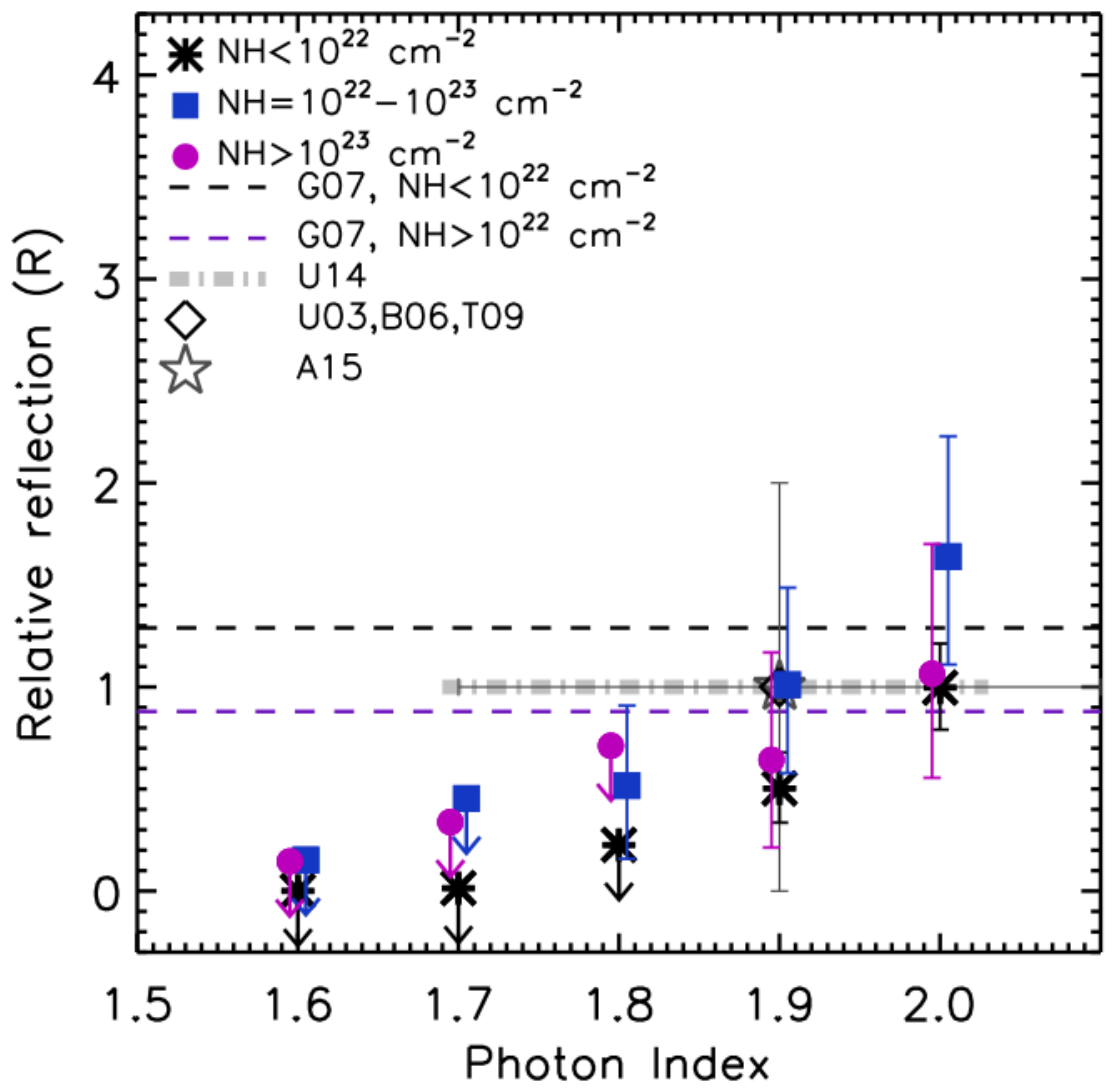}
\includegraphics[scale=0.67]{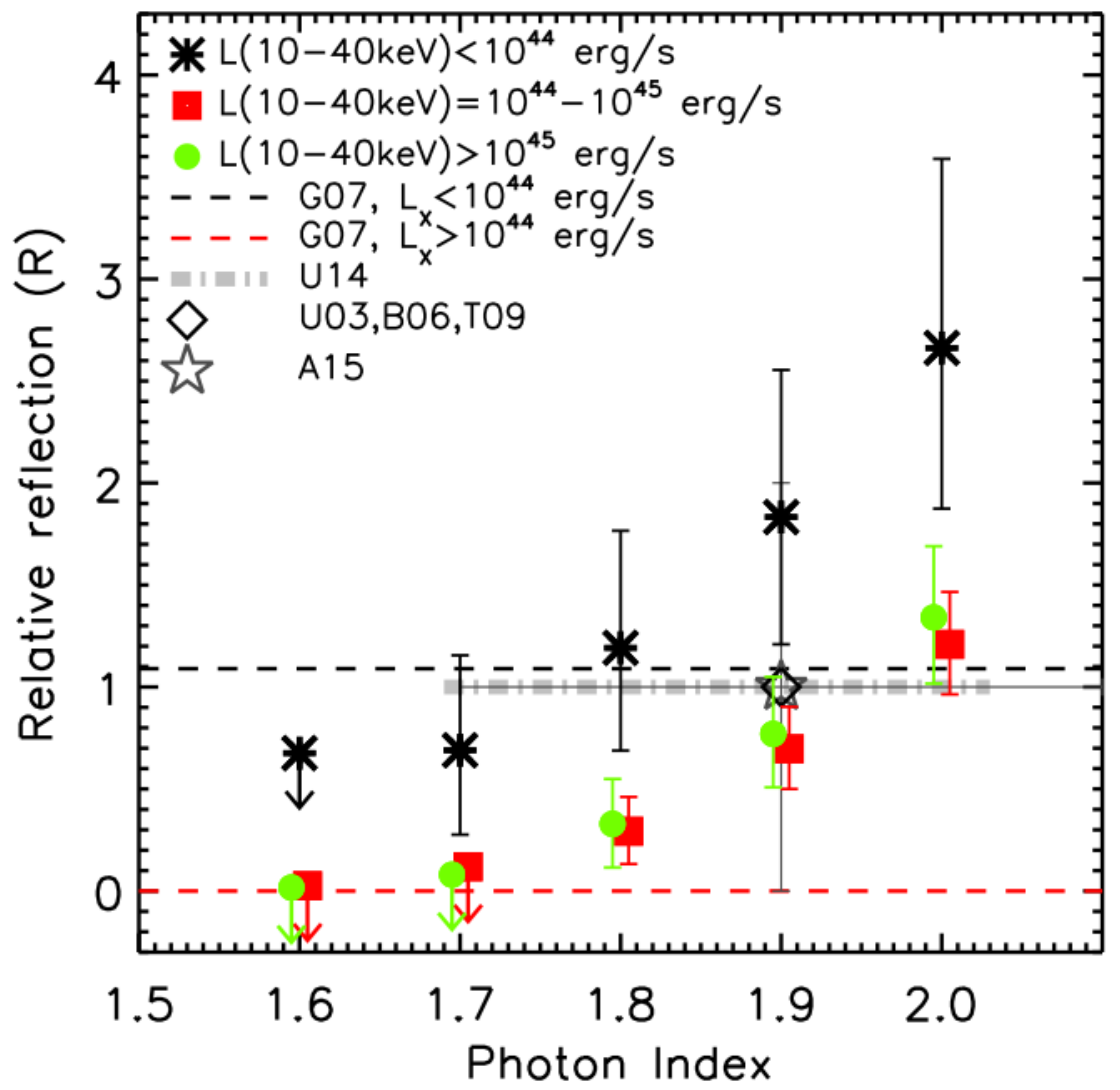}}
\caption{{\it Left}: Relative reflection strength ($R$) as a function of photon index ($\Gamma$) assumed in the model for the composite spectra in the three \nh\ bins: NH1 (black asterisks), NH2 (blue squares) and NH3 (magenta circles). The errors and upper limits are estimated at a 90\% confidence level. {\it Right}: $R$ versus $\Gamma$ in the three luminosity bins: L1 (black asterisks), L2 (red squares) and L3 (green circles). In the plots we indicate the AGN spectral model parameters assumed in various CXB population synthesis models and XLF studies: \citet{gilli2007} (G07; dashed lines); \citet{ueda2014} (U14; dot-dashed line); \citet{ueda2003}, \citet{ballantyne2006} and \citet{treister2009} (U03, B06 and T09, respectively; black diamond); \citet{aird2015b} (A15; grey star).}
\label{fig.refl}
\end{figure*}

To constrain the amount of reflection in these CT AGN candidates, we resort again to the {\tt pexrav} model with the addition of a soft-scattered component. The resulting power-law slope is $\Gamma={1.88_{-0.22}^{+1.13}}$ and the column density is \nh$=({1.5_{-0.6}^{+1.3}})\times10^{24}$~cm$^{-2}$, consistent with the CT regime; the reflection fraction, however, is not well constrained in these spectra, ${R<3.64}$ (${R<2.22}$ for fixed ${\Gamma=1.8}$).

Comparing these results with those obtained for the composite spectrum of the sources in the NH3 bin (which also includes typically the five sources analyzed here), there seems to be disagreement in the obtained parameters, especially the flattening trend of the photon index (see Fig. \ref{fig.gamnh} and Sect. \ref{nhbins}), since from the joint fitting we obtain steeper $\Gamma$ than those obtained from the NH3 composite (Fig. \ref{fig.rcl}). We note, however, that the spectral slopes are consistent within the uncertainties. Moreover, we need to account for the fact that the composite spectra in the NH3 bin are only fitted above 3.5~keV (rest frame), while for the joint fit we also include softer energy data. To allow for a fair comparison between the results, we tested our joint-fitting analysis limiting the data to the same energy range used in the composite spectra (and removing the soft power-law component from the model). The baseline model in this case yields a much flatter photon index of $\Gamma={0.72^{+0.54}_{-0.46}}$ and \nh${<2.3}\times10^{23}$~cm$^{-2}$. Similarly, performing a joint fit with the {\tt pexrav} model, in the same energy range covered by the composite spectra (i.e., rest frame $E\ge{3.5}$~keV for the NH3 bin), we obtained: $\Gamma={0.97_{-0.36}^{+0.52}}$, \nh$={<2.1}\times10^{23}$~cm$^{-2}$ and a reflection fraction of ${R<1.01}$ (${R>0.98}$ for $\Gamma=1.8$ fixed). Within the large uncertainties, these parameters could be broadly consistent with the values and obtained from the NH3 composite spectrum (Fig. \ref{fig.rcl}); however, for $\Gamma=1.8$ we only obtain a lower limit on $R$ for the CT AGN candidates.  
The poor constraints obtained for the parameters in these two spectral fits, suggest that the soft-scattered component is still needed for such high column densities, and might also partly explain the flat photon index obtained for the NH3 bin spectrum. We argue that when the broader energy range is used in the joint fitting analysis, the photon index is mainly constrained by the soft-energy component. This could bias the results, as the soft component can be produced by various processes, not necessarily related to the nuclear AGN emission (e.g., star formation in the host galaxy), and have steeper slope than the intrinsic power law. On the other hand, the small number of photons detected from the heavily-absorbed primary AGN emission leads to the degeneracy between $\Gamma$ and \nh, as described in Sect. \ref{blnl}.

\section{Compton reflection and the CXB}\label{cxb}

Several previous works have highlighted the problem of parameter degeneracies in the synthesis models of the CXB (e.g., \citealt{gandi2007, treister2009,akylas2012}) and on the model's assumptions for the XLF (e.g., \citealt{aird2015b}), which prevent us from deriving important constraints, such as the fraction of CT AGN contributing to the CXB and their space density. Amongst all the parameters the amount of Compton reflection assumed in the spectral models yields the largest uncertainty on the CT AGN population (\citealt{akylas2012}). 
Besides directly observing and identifying CT AGN in the available X-ray surveys and measuring directly the true \nh\ distribution of AGN at all redshifts, which has proved to be challenging even with the deepest data, the only solution to break these degeneracies is to better measure the intrinsic spectral properties of the AGN population, and in particular the Compton-reflection fraction. Such measurements have been performed in the local Universe using {\it Swift-BAT} and {\it INTEGRAL} observatories (\citealt{molina2009, burlon2011, ricci2011, vasudevan2013,ballantyne2014, esposito2016}), which are sensitive to the hard X-ray energies needed to directly probe the peak of the Compton-reflection hump ($E\approx20-30$~keV, rest frame). However, \nus\ is the only observatory available to date that allows such studies at higher redshifts ($z\approx1$), thanks to its higher sensitivity ($\sim$2 orders of magnitude) at $E>10$~keV  compared to previous observatories.

The measurements we obtained from our composite spectra for the full \nus\ AGN sample ($R\approx0.5$; see Sect. \ref{blnl} and Fig. \ref{fig.rcl}), for typical $\Gamma=1.8$, are lower than the Compton-reflection strength measured for sources in the local Universe. For instance, \citet{ballantyne2014}, who investigated the mean $0.5-200$~keV spectrum of local AGN, found an average reflection strength of $R=1.7^{+1.7}_{-0.9}$, consistent with several previous works (e.g., \citealt{molina2009, burlon2011,ricci2011, vasudevan2013}). It is important to note, however, that the samples investigated in the local universe are usually dominated by Seyfert galaxies (i.e., with $L_{\rm X}<10^{44}$~\ergs), while in our sample the vast majority of the sources ($\sim$70\%) are quasars ($L_{\rm X}>10^{44}$~\ergs) for which we constrain a much weaker reflection fraction than for the lower-luminosity AGN. On the other hand, the results we obtain for the $L_{\rm X}<10^{44}$~\ergs\ AGN (L1 bin) are in good agreement with the above mentioned studies in the local Universe, possibly indicating that there is no significant evolution of the source spectra, and therefore their intrinsic physical properties, between redshift $z\approx0$ and $z\approx1$ (the median redshift of our sample).

To compare our results with the typical assumptions made by several AGN synthesis models of the CXB (e.g., \citealt{ueda2003,ballantyne2006,treister2009, gilli2007,ueda2014}), we performed here several spectral fits with photon index fixed at various typical values ($\Gamma=1.8\pm0.2$, usually assumed in the CXB models) to obtain better constraints on the amount of Compton reflection in our composite spectra for each assumed spectral slope. This is to overcome the degeneracy we found between $\Gamma$ and $R$ (Sect. \ref{nhbins}). In Figure \ref{fig.refl} we show the constraints on $R$ as a function of the assumed spectral slope for each of the three \nh\ bins (left, see Sect. \ref{nhbins}) and for the three luminosity bins (right, see Sect. \ref{lbins}). In general, at all ${\Gamma}$, we do not find a significant difference in R between the three \nh\ bins (when high and low-luminosity sources are combined together); for relatively flat photon indices ${\Gamma=1.6-1.8}$, the reflection fraction tends to be below 1, while for steeper $\Gamma$ it {might reach} values up to $R\approx1.5-2.0$ for the obscured sources (NH2 and NH3), {considering the uncertainties,} and $R\approx1$ for the unobscured ones (NH1).

The same analysis in the three luminosity bins (Fig. \ref{fig.refl}, right) shows that at all photon indices, $R$ tends to be higher for the low-luminosity sources (L1) than for the high-luminosity ones (L2 and L3; see Sect. \ref{lbins}). For flat photon indices ($\Gamma\le1.7$), at high luminosities (L2 and L3), the reflection fraction is consistent with $R=0$, while for steeper photon indices ($\Gamma=1.8-2.0$) we find significant reflection fractions also for the high-luminosity sources $R\approx0.3-1.3$. At low luminosities (L1), the reflection fraction ranges from $R\approx0.7$ to ${R\approx2.7}$ (at steeper photon indices). 

Most of the CXB models adopt a photon index of $\Gamma=1.9$ and $R=1$ in their parameterisation of the intrinsic spectra for all AGN, with no difference in luminosity and/or \nh\ (e.g., \citealt{ueda2003, ballantyne2006, treister2009, ueda2014}). On the other hand, \citet{gilli2007} assume different reflection fractions for absorbed and unabsorbed AGN with $L_{\rm X}<10^{44}$~\ergs\ (${R=0.88}$ and $R=1.3$, respectively; following \citealt{comastri1995}), implying that the reflection is mainly due to scattered radiation from the accretion disc; for all quasars ($L_{\rm X}>10^{44}$~\ergs) they assume $R=0$.

Our findings are in line with the assumptions made in the \citet{gilli2007} model for the high- and low-luminosity sources, however, there are some differences{, which could potentially have an impact in the models}. For instance, \citet{gilli2007} assume a larger reflection fraction for unobscured AGN compared to the obscured AGN, while from our analyses, at low luminosities, the obscured sources seem to have larger $R$ values (see figure \ref{fig.ewR}, left). Moreover, although the reflection fraction we find for the high-luminosity sources is small ($R{\approx0.3}$ for ${\Gamma=1.8}$; see Fig. \ref{fig.refl}), it makes a contribution of $\sim${12}\% to the flux at 10-40~keV compared to a model with $R=0$.      
Conversely, a reflection fraction of $R=1$, as assumed by the above mentioned models for all AGN, is consistent with our values for the low-luminosity AGN (although higher $R$ values should also be allowed in the models, according to our results). However, it would overestimate the reflection contribution for the high-luminosity AGN, compared to our results, by $\approx10-20$\% in the $10-40$~keV flux. These differences might not have a big impact on the CXB models, as the majority of the contribution to the CXB spectrum comes from low-luminosity sources ($\sim$75\% from $L_{\rm X}<10^{44}$~\ergs\ AGN according to the \citealt{gilli2007} model).  
However, detailed modelling, accounting for all these differences, and folding in the AGN XLF, are necessary to reliably assess whether there are significant discrepancies between our results and the typical assumptions of the CXB synthesis models and how much our constraints on $R$ would impact on these models. For instance, higher (lower) values of $R$ in the CXB models would require a smaller (higher) fraction of CT AGN to reproduce the CXB spectrum peak at $E\approx20-30$~keV, and this would place important constraints on the overall accreting BH population in the Universe.

\citet{aird2015b} compared two different models of the XLF by \citet{aird2015a} and \citet{ueda2014} to reproduce the number of sources observed by \nus. The main difference between the models are in the assumed \nh\ distribution and on the fraction of CT AGN amongst the total AGN population. They find that both models can reproduce the observations, however, in the \citet{ueda2014} model a large contribution from Compton reflection ($R\approx2$) in the source spectra would be necessary to obtain good agreement with the observations (\citealt{aird2015b}). With our measurements of the average spectral properties of the \nus\ sources we find that the typical contribution from Compton reflection in the spectra is relatively small, and although values up to $R\approx2$ are found for the low-luminosity AGN (see Fig \ref{fig.refl}), these constitute a small fraction of the population probed by \nus, especially at $z>0.5$. Therefore, our results seem to support the \citet{aird2015a} model over the \citet{ueda2014} model, under the assumptions made in \citet{aird2015b}.

\section{Summary}

Constructing the rest-frame composite spectra for all the \nus\ detected AGN in the E-CDFS, EGS and COSMOS fields, using \ch\ and \nus\ data, we have investigated the average spectral properties of the BL and NL AGN as well as of the HB and SB-detected sources; we also studied the spectral properties as a function of X-ray absorption (\nh) and of X-ray luminosity at $10-40$~keV, producing and analysing the composite spectra in three different \nh\ bins and $L_{\rm 10-40~keV}$ bins. With this work we find: 
\begin{itemize}
\item[-] The average broad X-ray band ($2-40$~keV, rest frame) spectral slope for BL AGN and unabsorbed sources is $\Gamma\approx1.8$, consistent with previous results at similar energy ranges (\citealt{burlon2011, ricci2011,alexander2013}); while for NL AGN and X-ray absorbed sources we typically find flatter $\Gamma\approx{1.4-1.6}$, likely due to the effects of absorption and the contribution of Compton reflection at high energies (Sects. \ref{blnl} and \ref{nhbins}). 

\item[-] The average reflection fraction ($R$) found in our spectra is $R\approx0.5$ for typical $\Gamma=1.8$. Assuming the same intrinsic spectral slope for all the sources, to avoid parameter degeneracy, NL AGN and absorbed sources tend to have higher $R$ values ($R\approx0.5-0.7$) compared to the BL and unabsorbed AGN ($R\lesssim0.2${; see also Fig.\ref{fig.ewR}, left}). However, better counting statistics in the most heavily obscured AGN spectra are needed to assess whether there is any real correlation between $R$ and \nh\ (Sects. \ref{blnl} and \ref{nhbins}).
\item[-] We find that the reflection strength for low-luminosity AGN ($L_{\rm 10-40~keV}<10^{44}$~\ergs) is relatively high ${R\approx1.2}$ and decreases  at high luminosities ($L_{\rm 10-40~keV}>10^{44}$~\ergs), for which ${R\approx0.3}$ (Sect. \ref{lbins}), as found in some previous studies (e.g., \citealt{bianchi2007, liu2016}). Although we cannot establish a statistically significant correlation, due to the small number of data points an their uncertainties, this decreasing trend of $R$ with luminosity is seen at all assumed $\Gamma$ (Sect. \ref{cxb}), and seems to be present also dividing the high and low-luminosity sources into obscured and unobscured (see Sect. \ref{Rline}).

\item[-] We find that the EW of the iron K$\alpha$ line has a similar dependence with \nh\ and luminosity to that seen for the reflection strength, as it would be expected, since they are supposed to originate from the same reflecting material in the nuclear region. In particular the EW { seems to decrease} with the X-ray luminosity; { with the current data, the anti-correlation is not statistically significant, due to the large uncertainties, however, the trend is consistent with the X-ray Baldwin effect (see Sect. \ref{lbins}),} found in previous works (e.g., \citealt{iwasawa1993, nandra1997, pagek2004, bianchi2007, ricci2013a}).

\item[-] Our results are in line with the assumptions made in the \citet{gilli2007} CXB model for the intrinsic AGN spectra, however, there are some differences, also comparing to other CXB models (e.g., \citealt{ueda2003, ballantyne2006, treister2009, ueda2014}). Detailed modelling, with our improved $R$ estimates, are needed to reliably assess whether our results have a significant impact on the CXB model results, e.g., on the CT AGN fraction needed to reproduce the CXB peak (e.g., \citealt{treister2009,akylas2012}).  

\item[-] From the simple spectral fitting of individual sources we identify five CT AGN candidates, which have hydrogen column density values in the CT range, within the errors. The joint-spectral fitting of these sources with more complex, physical models (see Sect. \ref{ct}) provides solutions consistent with a CT interpretation. However, a strong Fe K$\alpha$ emission line, which is a typical feature in CT AGN (EW$\sim1$~keV), could not be constrained in the fit (EW${<0.7}$~keV from the joint-fitting results) {and it is not possible to confirm the CT nature of these sources individually, with the current data (except for \nus\ ID 330; see \citealt{civano2015})}.  
\end{itemize}
Further improvement on the constraints found in our work will likely be provided by the large AGN samples yielded by the \nus\  Serendipitous survey (\citealt{lansbury2017}). These sources have been excluded from our analyses due to the heterogeneity of the lower energy coverage, which would increase the systematic errors in our composite spectra. However, larger numbers of sources are needed to reduce the scatter on the composite spectra and therefore place tighter constraints on the average spectral parameters. Moreover, deeper \nus\ observations on the current survey fields would also be helpful to increase the S/N of the individual spectra; this would allow us to perform detailed analyses of the individual sources and thus provide important constraints on the distribution of the spectral parameters ($\Gamma$, \nh\ and $R$). 

\acknowledgments
We thank A. Corral for useful discussions about the spectral averaging method presented in her work. ADM thanks the financial support from the Max Plank Society and the UK Science and Technology Facilities Council (STFC, ST/L00075X/1, ADM and DMA; ST/K501979/1, GBL). FEB and ET acknowledge support from CONICYT-Chile (Basal-CATA PFB-06/2007, ``EMBIGGEN'' Anillo ACT1101, FONDECYT Regular 1141218 (FEB) and 1160999 (ET)); FEB also thanks the Ministry of Economy, Development, and Tourism's Millennium Science Initiative through grant IC120009, awarded to The Millennium Institute of Astrophysics, MAS. PG thanks the STFC for support (ST/J003697/2). WNB thanks the Caltech \nus\ subcontract 44A-1092750 and NASA ADP grant NNX10AC99G. AC and LZ acknowledges support from the ASI/INAF grant I/037/12/0Ð 011/13.

This work was supported under NASA Contract No. NNG08FD60C, and made use of data from the \nus\ mission, a project led by the California Institute of Technology, managed by the Jet Propulsion Laboratory, and funded by the National Aeronautics and Space Administration. We thank the \nus\ Operations, Software and Calibration teams for support with the execution and analysis of these observations. This research has made use of the \nus\ Data Analysis Software (NuSTARDAS) jointly developed by the ASI Science Data Center (ASDC, Italy) and the California Institute of Technology (USA).

\bibliographystyle{apj}
%\bibliography{/Users/agnese1/Documents/work/mypapers/allpapers}

\begin{appendix}
\section{Spectral simulations}\label{ap}
To test the results obtained from our spectral stacking procedure and to understand the effects and distortions that might be introduced at various stages of the process we performed extensive spectral simulations. These are essential to understand the intrinsic spectral properties derived from the average spectra. We therefore performed various tests, simulating \nus\ and \ch\ spectra with: i) a range of photon indices ($\Gamma=1.6-2.0$), ii) a fixed photon index ($\Gamma=1.8$) and various levels column densities, and running the same stacking procedure used for the real data.

\subsection{Combining spectra with different spectral slopes}

We initially simulated \nus\ and \ch\ spectra as a simple power law with $\Gamma=1.8$ fixed, using the response and ancillary files extracted from our survey data, and a flux distribution similar to that of the real sources. For practicality, since the simulations are very time consuming, we only used the files of the sources detected in the E-CDFS field to simulate the spectra (i.e. 45 sources). When more spectra are needed to increase the counting statistics in the composite spectra (see below), we generated multiple simulated spectra for each real source (using photon randomization). We simulated both background and source spectra to test more closely the results from the real data. We applied to the simulated spectra the same procedure used for the real data (see Sect. \ref{stacks}), i.e., we fitted the individual background-subtracted spectra with a power-law model (with $\Gamma$ free to vary) and saved them in physical units (unfolded spectrum in XSPEC), we shifted the spectra to the rest frame, using the redshifts of our real sources, created a new energy grid for all the spectra, normalized them to the flux at $8-15$~keV and redistributed the fluxes in the new energy bins. Our average spectra are obtained by taking the median flux in each new energy bin, performing a {resampling} analysis to estimate the 1$\sigma$ errors. We then fitted the obtained spectrum with a power-law model to verify whether we can recover the input photon index of the simulated spectra. We obtained $\Gamma=1.77\pm0.03$, which is slightly lower than the input spectral slope, but in good agreement within the errors, with the input parameters. 
\begin{figure}
\centerline{
\includegraphics[scale=0.7]{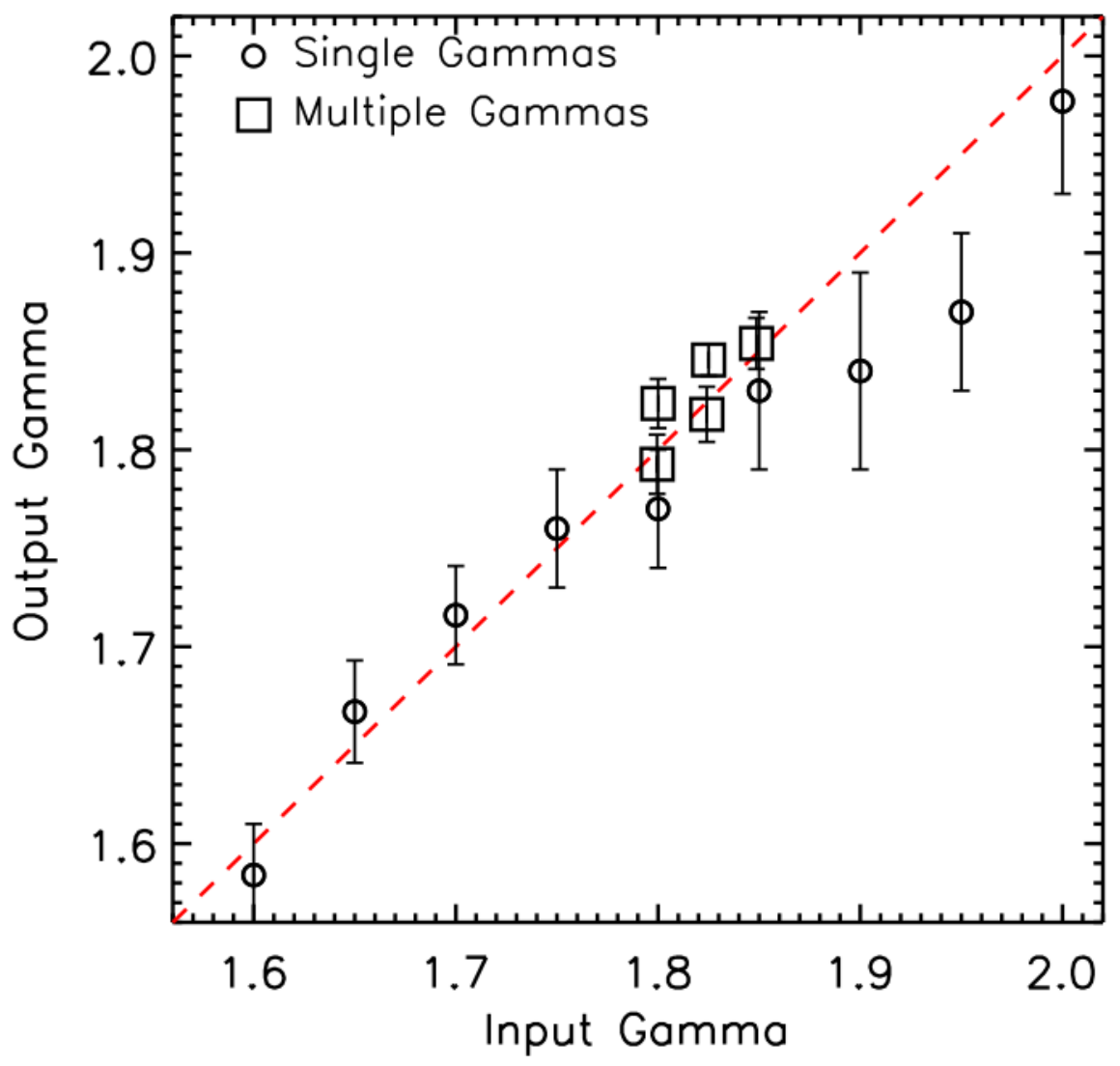}}
%\vspace{-0.7cm}
\caption{Spectral slopes (output $\Gamma$) obtained from the composite spectra produced using simulated \nus$+$\ch\ spectra with different spectral slopes (input $\Gamma$) in the range between 1.6 and 2.0. The circles represent the results from the composite spectra obtained combining spectra with one single value of $\Gamma$ (typically $\sim$45 simulated spectra for each composite), while the squares represent the results obtained combining spectra with different slopes (typically $>90$ simulated spectra for each composite); in these cases we plot the median of the input $\Gamma$ distribution on the x axis. The red dashed line represents the 1:1 line.} 
\label{fig.sim1}
\end{figure}

We performed the same test simulating spectra with different photon indices (45 spectra for each $\Gamma$ value), in the range of $\Gamma=1.6-2.0$, with a step-size of $\Delta \Gamma=0.5$ and obtaining the average spectrum for each input spectral slope. Subsequently, we also combined spectra with different slopes, to test whether we can recover the median input $\Gamma$ from our composite spectra or whether significant distortions are affecting the spectra. In Figure \ref{fig.sim1} we show the results of these tests by plotting the resulting spectral slopes, obtained by fitting the composite spectra with a power-law model, versus the input $\Gamma$ used to simulate the individual spectra. When spectra with different slopes are combined, we plot their median $\Gamma$ on { the} x axis. In general, the spectral slopes obtained from the composite spectra are in good agreement with the input $\Gamma$ of the simulated spectra used to produce the composites. However, for steep slopes ($\Gamma>1.9$), the resulting $\Gamma$ tends to be flatter than the input values. This is likely because when the spectra are steep, there are fewer counts contributing to the spectra at high energies (e.g., in the \nus\ band), increasing the scatter in the composite spectra at $E\gtrsim10$~keV, thus impairing the constraints on the intrinsic spectral slope. This results in a slightly flatter value of $\Gamma$. We note  that this issue is affecting the results only when a relatively small number of spectra are used to produce the composites (for instance, 45 spectra in our tests with single $\Gamma$ values), while when we increase the number of spectra (hence, the counting statistics), as in the case of the composites obtained from spectra with different $\Gamma$ values, the issue is no longer present. 

We can therefore conclude that our averaging method does not introduce significant distortions to the final composite spectra when combining spectra with different slopes, in the case where a simple power-law model is assumed. From our tests on these simulated composite spectra we can reliably recover the input $\Gamma$ of the individual spectra, and/or their median value, when a distribution of $\Gamma$ is assumed.
 
\begin{figure}
\centerline{
\includegraphics[scale=0.38]{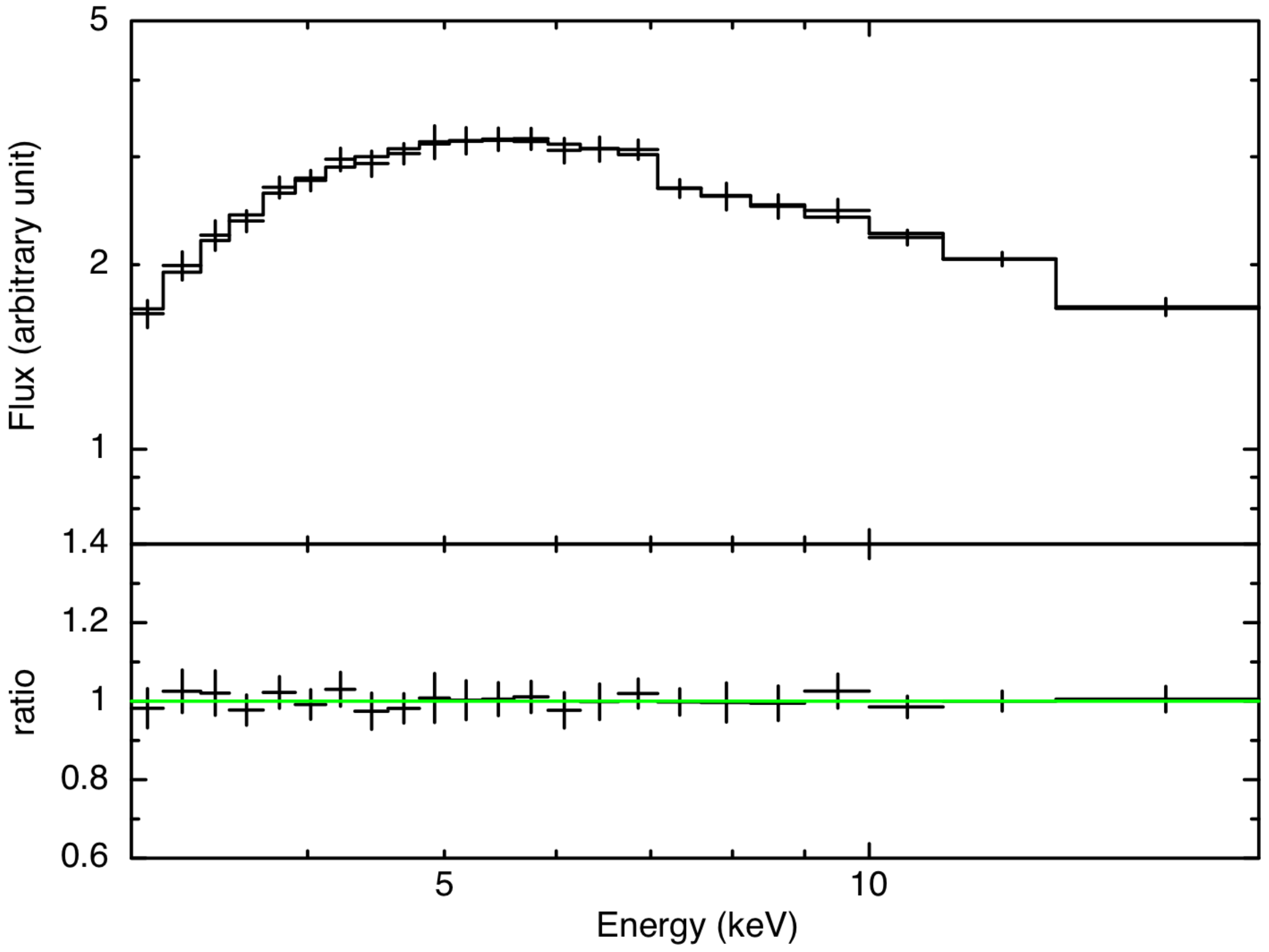}}
\vspace{-0.2cm}
\caption{Median rest-frame spectrum from the simulated \ch\ and \nus\ spectra with $\Gamma=1.8$ and \nh$=10^{23}$~cm$^{-2}$. To test the reliability of our averaging method and constrain the underlying continuum, the median spectrum was fitted with an absorbed power-law model between 3 and 30~keV: we recovered the input photon index $\Gamma=1.8$ and \nh$=10^{23}$~cm$^{-2}$ with a scatter of $\sim$1\% and $\sim$2\% respectively. The bottom panel shows the ratio between the simulated data and the model.}
\label{fig.sim2}
\end{figure}

\subsection{Simulations with various column densities: the effects of absorption}

An important effect that can create distortions in the composite spectra and modify the results from our analyses is the X-ray column density. 
To test how the X-ray absorption modifies the composite spectra, we simulated the \ch\ and \nus\ source spectra (and the relative background) with different values of \nh\ (\nh$=10^{21}, 10^{22}, 10^{23}, 5\times10^{23}$~cm$^{-2}$) and a fixed $\Gamma=1.8$, and produce the composite rest-frame spectra for individual values of \nh. An example is shown in Fig. \ref{fig.sim2}. Analysing the spectra (at $E\approx3-30$~keV, rest frame) in all cases we obtained parameters in good agreement with the input values; the results are summarised in Table \ref{tab.a1}. We note that for the \nh$=10^{21}$~cm$^{-2}$ composite, where we obtained a much lower \nh\ value than the input simulated spectra, the discrepancy is due to the fact that we cannot constrain such a low value of \nh\ from our composite spectrum as it spans the energy range $E\approx3-40$~keV, rest frame. For high levels of \nh\ the resulting photon index tends to be slightly flatter than the input $\Gamma=1.8$, but it is always consistent within the errors. In these cases, the flattening is likely due to a decrease of the number of counts in the composite spectra at low energies, due to the absorption, thus yielding poorer constraints on the intrinsic spectral slope. 

\begin{table}
\caption{Spectral parameters of the simulated composite spectra, obtained combining simulated \ch$+$\nus\ spectra with $\Gamma=1.8$ and different levels of X-ray absorption.}
\begin{center}
\begin{tabular}{c c c c}
\hline
\rule[-1.5mm]{0pt}{4ex} Input \nh$^{a}$ & No. of sources & $\Gamma$ & \nh$^{a}$ \\
\hline
\rule[-1.5mm]{0pt}{4ex}0.1              & 45       & $1.83\pm0.01$ & $<0.05$    \\ 
\rule[-1.5mm]{0pt}{4ex}1.0              & 45       & $1.82\pm0.02$ & $0.9\pm0.1$  \\ 
\rule[-1.5mm]{0pt}{4ex}10.0             & 45       & $1.78\pm0.02$ & $10.1\pm0.2$ \\
\rule[-1.5mm]{0pt}{4ex}50.0             & 45       & $1.75\pm0.07$ & $51.4\pm2.7$ \\ 
\rule[-1.5mm]{0pt}{4ex}0.1,1.0  (a)     & 90       & $1.79\pm0.03$ & $0.6\pm0.1$  \\ 
\rule[-1.5mm]{0pt}{4ex}1.0,10.0 (b)     & 90 (180) & $1.72\pm0.05$ ($1.78\pm0.03$) & $6.3\pm0.7$ ($6.3\pm0.4$)\\
\rule[-1.5mm]{0pt}{4ex}0.1,1.0,10.0 (c) & 135      & $1.76\pm0.04$ & $4.0\pm0.7$ \\
\hline
\hline
\end{tabular}
\end{center}
Notes: $^a$ The hydrogen column density is expressed in units of $10^{22}$~cm$^{-2}$. 
\label{tab.a1}
\end{table}%

We then produced composite spectra combining simulated source spectra with different levels of absorption, to see whether, also in these cases, the resulting spectral parameters are in agreement with the input distributions. We note that we do not expect to obtain the true median value of \nh\ from the composite spectra as the absorption is not a linear parameter. However, the resulting \nh\ should be somewhat consistent with (or within the range of) the input values. We therefore combined spectra with: a) \nh$=10^{21}, 10^{22}$~cm$^{-2}$; b) \nh$=10^{22}, 10^{23}$~cm$^{-2}$, and c) \nh$=10^{21}, 10^{22}, 10^{23}$~cm$^{-2}$. The spectral parameters obtained from these composite spectra are reported in Table \ref{tab.a1}. For the unabsorbed and lightly absorbed sources (composite spectrum ``a'') the resulting parameters are in very good agreement with those of the input simulated spectra, while for the spectra ``b'' and ``c'', which also include more absorbed sources, the resulting spectral slope tends to be flatter than those of the input sources. The flattening is mainly due, as described above, to a decrease of the number of counts in the composite spectra for the absorbed sources. Indeed, we tested that increasing the number of input spectra to produce the composite ``b'' (e.g., from 90 to 180; see Table \ref{tab.a1}), thus increasing the counting statistics in the composite spectrum, we can recover a photon index in good agreement with the $\Gamma=1.8$ of the input spectra. However, another effect plays a non-negligible role in flattening the photon index of the composite spectra: combining spectra with different values of column densities, and therefore different photoelectric cut-off energy, can lead to a slight flattening in the spectral slope, as the typical absorption features get ``smoothed'' in the final composites, yielding a flatter $\Gamma$ and possibly an underestimation of \nh\ (since it is not possible to recover the true median value of \nh\ from the composite spectra). This effect is also seen in the real data (Sects. \ref{blnl} and \ref{nhbins} and Fig. \ref{fig.gamnh}). However, our simulations indicate that this ``artificial'' flattening of the spectra due to the averaging process, is relatively small ($\Delta \Gamma\approx0.2-0.8$) and cannot be the solely cause of the flat spectral slopes seen in the composites of the NL AGN and the heavily-absorbed sources (NH3), for which the resulting photon indices are $\Gamma\approx1.6$.  

\subsection{Arithmetic mean, geometric mean and median}
\begin{figure}
\centerline{
\includegraphics[scale=0.5]{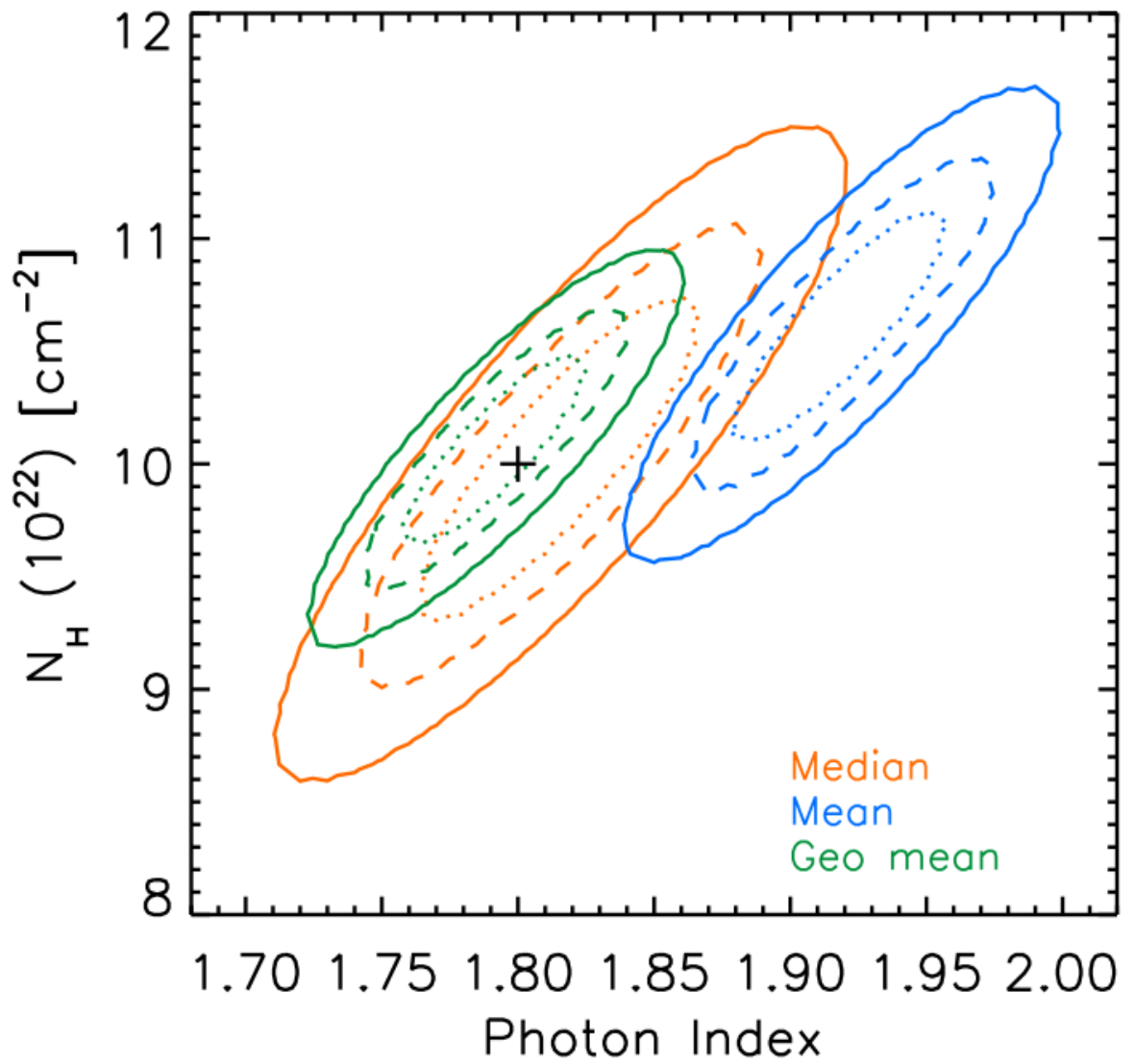}
\includegraphics[scale=0.5]{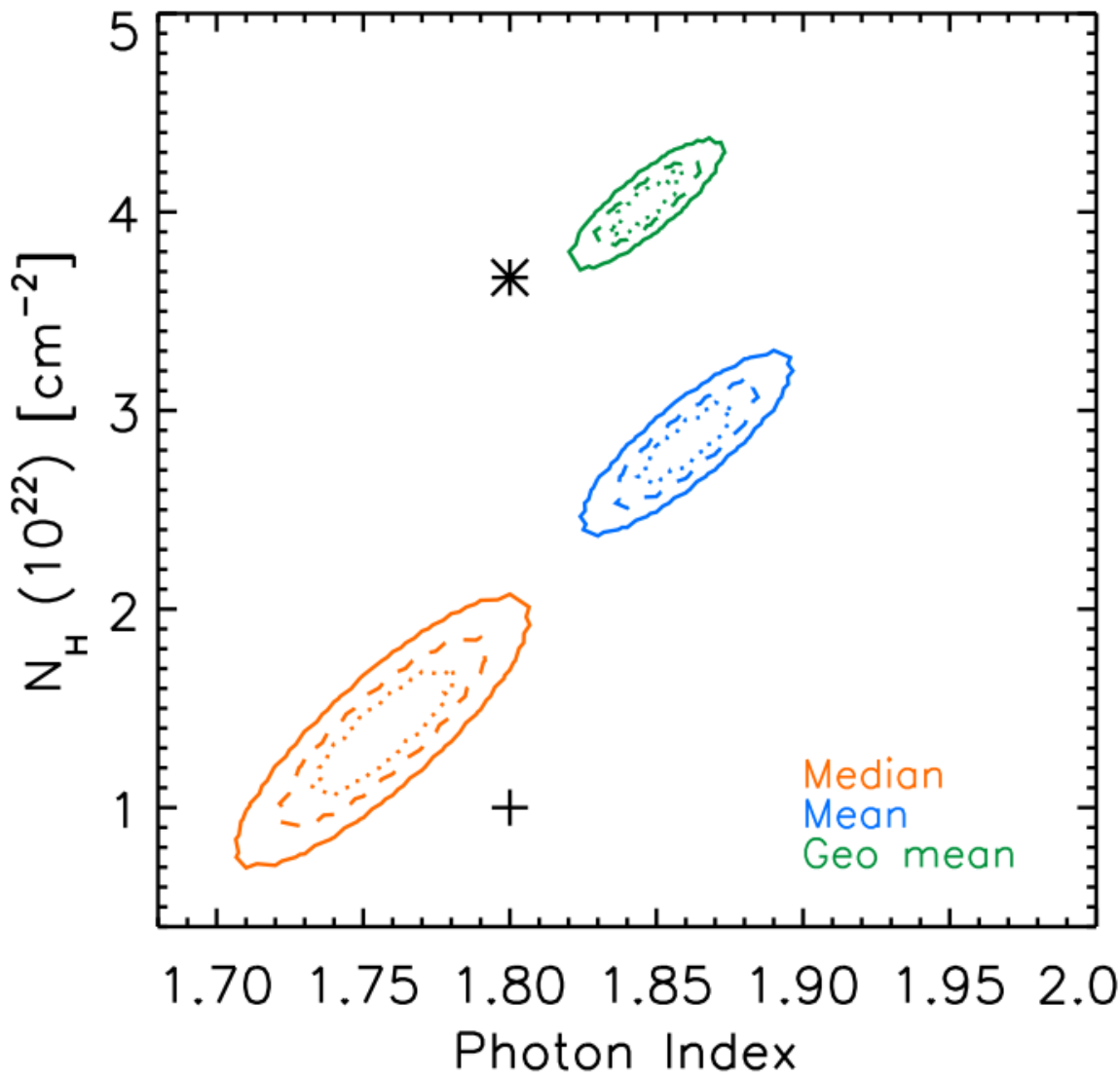}}

\caption{Confidence contours for the spectral parameters $\Gamma$ and \nh\ recovered from the composite simulated spectra, with input $\Gamma=1.8$ and \nh$=10^{23}$~cm$^{-2}$ (left) and $\Gamma=1.8$ and \nh$=0,10^{22},10^{23}$~cm$^{-2}$ (right), using the arithmetic mean (blue), median ({ orange}) and geometric mean (green) as averaging methods. The contours correspond to 68\%, 90\% and 99\% confidence levels { (dotted, dashed and solid curves, respectively)}. The black cross marks the median of the input parameters, and the asterisk marks the mean values.}
\label{fig.cont}
\end{figure}

There are different methods one can use to determine the average spectrum of a population. The arithmetic mean (or median), tend to preserve the relative fluxes of the emission features, while the geometric mean tends to preserve the global continuum shape, when it can be approximated by a power law (\citealt{vandenberk2001}). Since the aim of our analyses is to study the intrinsic AGN continuum, the geometric mean would be the most appropriate choice to derive our composite spectra. However, since the \nus\ spectra have typically high background levels, several energy bins in the individual spectra result in a null (or negative) flux value after background subtraction, due to the background fluctuations. These energy bins have to be excluded from the geometric mean, therefore biasing the resulting average flux. 

We performed several tests to assess the differences between these averaging methods and evaluate the best approach to use for our data. We initially used the simulated spectra (see details in the previous section) with known input spectral slopes and analyzed the resulting composite spectra. When we produce composite spectra for unabsorbed $\Gamma=1.8$ simulated spectra, the three averaging methods yield comparable results, recovering the input $\Gamma=1.8$. When we introduce various levels of absorption in our simulated spectra, however, the resulting composite spectra are affected by some distortions (see details below), and the averaging method used to produce the composite spectra yield different results. In Figure \ref{fig.cont} we show the confidence contours for the spectral parameters $\Gamma$ and \nh\ derived from the composite spectra. On the left the input spectra used for the composite are all simulated with the same $\Gamma=1.8$ and 
\nh$=10^{23}$~cm$^{-2}$ (which are the values we want to recover from the resulting composite spectra); on the right the input spectra used for the composite have a photon index $\Gamma=1.8$ and \nh$=0,10^{22},10^{23}$~cm$^{-2}$ in equal numbers. 
The median and the geometric mean provide comparable results in the first case, while the arithmetic mean yields a steeper photon index than the input value. When spectra with various column densities are stacked together to produce the composite, the parameters derived from the median spectrum are consistent with the input \nh\ values of the individual spectra (although, we do not expect to recover the true median \nh, as described in the previous section), whilst slightly underestimating the photon index, which is flatter than the input value $\Gamma=1.8$, but still consistent within the errors. On the other hand, $\Gamma$ tends to be overestimated in the composite generated adopting the geometric mean, as well as the arithmetic mean. We also note that, since the median is less sensitive to outliers or extreme values, the composite spectrum produced with this method is less noisy than those produced using the arithmetic or geometric means. We therefore favour the median as the method to produce our composite spectra (see also \citealt{falocco2012}), keeping in mind that the flattening of the photon index seen in the composite spectra of our real data is partially due to the stacking method (see previous section). 

\end{appendix}

\end{document}